\documentclass[entropy,article,accept,oneauthor,pdftex,10pt,a4paper]{mdpi} 
\graphicspath{ {Figures/}}

\firstpage{1} 
\articlenumber{567}
\doinum{10.3390/e19100567
}
\pubvolume{19}
\pubyear{2017}
\copyrightyear{2017}
\history{Received: 12 September 2017; Accepted: 23 October 2017; Published: {24 October 2017}}

\usepackage{upgreek}
\usepackage{amsmath}
\usepackage{amsfonts}
\usepackage{graphicx}
\usepackage{caption}
\usepackage[labelformat=simple]{subcaption}

\DeclareCaptionLabelFormat{subcaptionlabel}{\normalfont(\textbf{#2}\normalfont)}
\graphicspath{ {Figures/}} 
 \usepackage{amssymb}

\usepackage{epstopdf}
\usepackage{epsfig}
\usepackage{hyperref}

\usepackage{color}

\Title{Radially Excited AdS$_5$ Black Holes in Einstein--Maxwell--Chern--Simons Theory}

\Author{Jose Luis Bl\'azquez-Salcedo \orcidA{}}
 
\AuthorNames{Jose Luis Bl\'azquez-Salcedo}

\address[1]{%
 Institut f\"ur  Physik, Universit\"at Oldenburg, Postfach 2503, D-26111 Oldenburg, Germany; jose.blazquez.salcedo@uni-oldenburg.de; Tel.: +49-441-798-3469\\
 }

\abstract{
In the large coupling regime of the 5-dimensional Einstein--Maxwell--Chern--Simons theory, charged and rotating cohomogeneity-1 black holes form sequences of extremal and non-extremal radially excited configurations. These asymptotically global Anti-de Sitter (AdS$_5$) black holes form a discrete set of solutions, characterised by the vanishing of the total angular momenta, or the horizon angular velocity. However, the solutions are not static. In this paper, we study the branch structure that contains these excited states, and its relation with the static Reissner--Nordstr\"om-AdS black hole. Thermodynamic properties of these solutions are considered, revealing that the branches with lower excitation number can become thermodynamically unstable beyond certain critical solutions that depend on the free parameters of the configuration.
}


\begin{document}

%%%%%%%%%%%%%%%%%%%%%%%%%%%%%%%%%%%%%%%%%%%%%%%%%%%%%%%%%%%%%%%%%%%%
\section{Introduction}
%%%%%%%%%%%%%%%%%%%%%%%%%%%%%%%%%%%%%%%%%%%%%%%%%%%%%%%%%%%%%%%%%%%% 

Asymptotically Anti-de Sitter (AdS) black holes have gained a lot of interest since the AdS/Conformal Field Theory (CFT)
correspondence was proposed \cite{Witten:1998qj,Maldacena:1997re}. The correspondence suggests that fields propagating in asymptotically AdS$_D$ spaces would correspond to fields propagating in a ($D-1$)
 CFT theory in the AdS boundary. The~analysis of black holes in theories with negative cosmological constant could be used as a tool to describe non-perturbative states of the corresponding CFT theory. Of special interest are black holes in $D=5$ dimensions, since 
their AdS conformal boundary is a 4-dimensional space-time.

Many studies on asymptotically AdS$_5$ black holes can be found in the literature. The generalization of the Kerr black holes to an asymptotically AdS$_5$ solution was done in \cite{Hawking:1998kw}, which we will refer to as Myers--Perry-AdS (MP-AdS) black hole. This solution possesses two independent angular momenta and an event horizon of spherical topology. On the other hand, we have the generalization of the Reissner--Nordstr\"om black hole to an asymptotically AdS$_5$ solution, which has been studied in \cite{Mitra:1999ge,Chamblin:1999tk,Chamblin:1999hg}. We will refer to this black hole as the Reissner--Nordstr\"om-AdS (RN-AdS) black hole, and it is a static solution of the Einstein--Maxwell (EM) theory. It has been recently found that in odd dimensions the RN-AdS black hole is not the only static solution the theory allows with a non-trivial $U(1)$ gauge field, and, for instance, EM theory also contains purely magnetic static AdS black holes and solitons \cite{Blazquez-Salcedo:2016vwa}.

It is interesting to note that the rotating generalization of the RN-AdS black hole (i.e., the AdS$_5$ equivalent of the Kerr--Newmann black hole) is not known analytically, and has been investigated numerically for equal magnitude angular momenta in the non-extremal \cite{Kunz:2007jq} and extremal cases \cite{Blazquez-Salcedo:2016rkj}. In~addition, asymptotically AdS black holes in $D>4$ can have an event horizon topology different from spherical, e.g., black ring solutions \cite{Figueras:2014dta}.
  
Nevertheless, there are examples of analytical solutions of AdS black holes with rotation and electrically charged, the first found by Cvetic, Lu and Pope \cite{Cvetic:2004hs} (CLP black hole). This black hole is a~solution of the Einstein--Maxwell--Chern--Simons (EMCS) theory with negative cosmological constant, in the particular case in which the Chern--Simons (CS) coupling is fixed to $\lambda=\lambda_{SG}=1$. This theory describes the bosonic sector of the 5-dimensional minimally gauged supergravity (SUGRA). The black hole possesses equal-magnitude angular momenta, electric charge, and the event horizon has spherical topology. In the extremal limit, it is known that the solution contains a subset of black holes that preserve a fraction of supersymmetry \cite{Gutowski:2004ez}. 
In addition, the CLP black hole has been generalized to the case of unequal angular momenta, and to solutions of a more general SUGRA with additional field content \cite{Chong:2005hr,Chong:2006zx,Chong:2005da,Cvetic:2004ny} and asymptotic behavior \cite{Blazquez-Salcedo:2017cqm}. The thermodynamic properties of the CLP black hole with unequal angular momenta were studied in \cite{Grunau:2015ana}.

By making use of numerical techniques, it was recently found that the CLP black hole can be generalized to non-SUGRA EMCS theories with a generic CS coupling \cite{Blazquez-Salcedo:2016rkj}. Similar configurations were studied using a perturbative approach \cite{Mir:2016dio}. 
These black holes are similar to the CLP solution, in~the sense that they possess equal-magnitude angular momenta and spherical topology of the event horizon. 
However, the branch structure of the solutions as well as several properties of the black holes depend on the value of the CS coupling. 
In particular, the study of the properties of the black holes with CS coupling between $\lambda=0$ and $\lambda=2\lambda_{SG}$ realized in \cite{Blazquez-Salcedo:2016rkj} 
showed that new features arise when the CS coupling is small enough. For example, there are two disconnected branches of extremal black holes, meaning that the extremal RN-AdS solution cannot be continuously connected with the extremal MP-AdS black hole. However, for intermediate values of the coupling, the connection between both branches is restablished, resulting in sets of solutions with properties that can be approximated (qualitatively) by 
the properties of the CPL black hole.

The case for large CS coupling was considered before in the asymptotically flat case, where it was shown that when $\lambda > 2\lambda_{SG}$, a complicated branch structure of solutions surrounding the static black hole exists \cite{Blazquez-Salcedo:2013muz,Blazquez-Salcedo:2015kja}. The case for large CS coupling with the AdS$_5$ asymptotics was briefly introduced in \cite{Blazquez-Salcedo:2016rkj} too,
where it was mentioned that similar features to the asymptotically flat case can also be found.
It is the purpose of this paper to study in a systematic way the space of solutions in the large CS coupling regime with the AdS$_5$ asymptotics, 
a study, which, to the best of our knowledge, has not been done before.
We will show that, even though the asymptotics are different, most of the qualitative features from the asymptotically flat configurations are still present after the introduction of a negative cosmological constant. 
For instance, when $\lambda > 2\lambda_{SG},$ once again, new features of the theory arise. 
For~example,
we will see that such black holes possess non-uniqueness with respect global and horizon quantities, and contains sequences of extremal radially excited non-static black holes with vanishing total angular momentum or angular velocity. Hence, these new solutions share many properties with the asymptotically flat ones, but the new asymptotic behaviour makes them of interest in the context of the AdS/CFT correspondence.

We will consider also the relation of these new excited AdS extremal black holes with the non extremal RN-AdS solution. 
For instance, we will see that the extremal RN-AdS solution becomes isolated from the rest of extremal rotating and charged black holes. 
Similarly to what happens in the small CS coupling regime, the extremal MP-AdS black hole cannot be continuosly connected with the extremal RN-AdS configuration \cite{Blazquez-Salcedo:2016rkj}.
However, instead of two disconnected branches, here one finds infinitely many branches between the MP-AdS and the RN-AdS black holes. Inside these branches, two~sets of non-static radially excited configurations (one with vanishing total angular momentum, and the other one with vanishing angular velocity) are found, characterised by an integer excitation number. They can be continuously connected with the non-extremal RN-AdS black hole if their temperature is increased beyond some critical point. These new features are not found when $\lambda<2\lambda_{SG},$ and, in~particular, they~are not present in the CPL solution.
Since we lack the knowledge of an analytical solution describing these black holes, we use numerical techniques to generate them and analyze their~charges.\enlargethispage{0.5cm}

In Section \ref{s2}, we describe the general framework, Ansatz, charges and parameters characterizing these black holes. In Section \ref{s3}, we present the numerical results and study the branch structure of the solutions, and properties of the different classes of radially excited black holes. In Section \ref{s4}, we finish with conclusions and an outlook.

Let us start by reviewing briefly the theory and describing the Ansatz we use to study these~configurations.

%%%%%%%%%%%%%%%%%%%%%%%%%%%%%%%%%%%%%%%%%%
\section{AdS$_5$ Black Holes in Einstein--Maxwell--Chern--Simons Theory} \label{s2}
\vspace{-6pt}

\subsection{The Theory}

The action of the Einstein--Maxwell--Chern--Simons theory in five dimensions with a negative cosmological constant can be written as

\begin{equation} 
\label{action}
I= -\frac{1}{16\uppi G_5} \int_{{\cal M}} d^5x \sqrt{-g} \biggl[ 
R +\frac{12}{L^2}
-F_{\mu \nu} F^{\mu \nu} 
+
\frac{2\lambda}{3\sqrt{3}}\varepsilon^{\mu\nu\alpha\beta\gamma}A_{\mu}F_{\nu\alpha}F_{\beta\gamma}
 \biggr ],
\end{equation}
where the determinant of the metric is $g$ and $R$ is the curvature scalar. The theory contains a $U(1)$ gauge potential,
$A$, with the field strength tensor defined as 
$ F_{\mu \nu} = \partial_\mu A_\nu -\partial_\nu A_\mu $. 
The last piece of the action is the CS term, and $\varepsilon^{\mu\nu\alpha\beta\gamma}$ is the totally antisymmetric tensor.
Note that we have introduced a~negative cosmological constant, $\Lambda=-\frac{12}{L^2}$, where $L$ is the AdS length scale.
$G_5$ is Newton's constant in five dimensions.  

This action is a generalization of the minimally coupled supergravity action, where we allow for an arbitrary coupling of the Chern--Simons term $\lambda$. This arbitrary coupling is not trivial, and the theories are not equivalent. Minimally gauged supergravity is only recovered when $\lambda=\lambda_{SG}=1$.

From the action, we can derive the field equations of the theory. We have the Einstein equations
\begin{equation}
\label{Einstein_equation}
G_{\mu\nu} + \Lambda g_{\mu\nu} =2\left(F_{\mu\rho}{F^{\rho}}_{\nu}-\frac{1}{4}F^2 g_{\mu\nu}\right),
\end{equation}
and the generalized Maxwell equations
\begin{equation}
\label{Maxwell_equation}
\nabla_{\nu} F^{\mu\nu} + \frac{\lambda}{2\sqrt{3}}\varepsilon^{\mu\nu\alpha\beta\gamma}F_{\nu\alpha}F_{\beta\gamma}=0.
\end{equation}

Note that the Chern--Simons term does not modify the definition of the electro-magnetic stress-energy tensor, and hence the Einstein Equations (\ref{Einstein_equation}) are equivalent to those of pure EM-AdS theory. However, the CS interaction introduces a nonlinear term into the Maxwell Equation (\ref{Maxwell_equation}), where~the coupling constant $\lambda$ can be understood as a parameter controlling the strength of the nonlinearity. As noted before, in this paper, we are interested in the large $\lambda$ case, with $\lambda > 2$ always.

\subsection{Ansatz, Asymptotics and Charges}

The black holes we study are directly related to the $5D$ generalization of the Kerr--Newmann-AdS black hole, and they satisfy similar requirements. First, we want the space-time to be stationary and axi-symmetric. Hence, we can define an asymptotically time-like vector field $\xi = \partial_t$, together with two periodic space-like vector fields $\eta_{(k)}=\partial_{\varphi_ {(k)}}$, where $(k)=1,2$. We will limit ourselves to topologically spherical event horizons. Additionally, we are interested in cohomogeneity-1 black holes, with both angular momenta of equal magnitude (note the isometry group of the space-time is $R \times U(2)$). 

By choosing coordinates  $t$, $r$, $\theta$, $\varphi_1$ and $\varphi_2$ adapted to these symmetries, we can parametrize the black holes with the following Ansatz for the metric:

\begin{equation}\label{metric}
\begin{array}{ll}
ds^2 =& -b(r) dt^2 + \frac{1}{u(r)}dr^2 + g(r)d\theta^2  
+ p(r)\sin^2\theta \left( d \varphi_1 -\frac{\omega(r)}{r}dt
\right)^2 \\  
& + p(r)\cos^2\theta \left( d \varphi_2
  -\frac{\omega(r)}{r}dt \right)^2
+ (g(r)-p(r))\sin^2\theta \cos^2\theta
   \left( d \varphi_1  - d \varphi_2\right)^2,
   \end{array}
\end{equation}
and for the gauge field:
\begin{equation}
\label{gfield}
A_\mu dx^\mu  = a_0(r) dt 
+ a_{\varphi}(r) (\sin^2 \theta d\varphi_1
  +\cos^2 \theta d\varphi_2),
\end{equation}
where $\theta \in [0,\pi/2]$,
$\varphi_{(k)} \in [0,2\pi]$.
Because of the asymptotical AdS$_5$ structure of the space-time, it is convenient to choose the metric functions to satisfy these relations 
\begin{eqnarray}
u(r) &=& \frac{f(r)}{m(r)}\left(\frac{r^2}{L^2}+1\right), \, \,
b(r) = f(r)\left(\frac{r^2}{L^2}+1\right), 
 \, \,
g(r) = \frac{m(r)}{f(r)}r^2, \, \,
p(r) = \frac{n(r)}{f(r)}r^2.
\label{param}
\end{eqnarray}

The main advantage of the cohomogeneity-1 symmetry is that all the ignorance we have of the solutions is parametrized by the unknown functions $f(r)$, $m(r)$, $n(r)$ and $\omega(r)$ for the metric, and $a_0(r)$ and $a_{\varphi}(r)$ for the gauge field. These functions only depend on the radial coordinate, and hence the field equations reduce to a system of nonlinear ordinary differential equations.

This radial coordinate is a quasi-isotropic coordinate, and we will consider solutions only in the domain $r \in [r_H,\infty)$, where $r=r_H$ is the exterior event horizon. Surfaces of constant $t, r$ are squashed $S^3$ hypersurfaces. 

As $r {\rightarrow} \infty$, the metric approaches AdS$_5$ and the functions become $f(\infty)=1$, $m(\infty)=1$, $n(\infty)=1$ and $\omega(\infty)=0$. A series expansion reveals that the next-to-leading terms of the Ansatz functions behave like:
\begin{eqnarray}
f(r) - 1 &\sim & \frac{\alpha}{r^4}, \ \ 
m(r) - 1 \sim  \frac{\beta}{r^4}, \nonumber \\
n(r) - 1 &\sim & \frac{3(\alpha-\beta)}{r^4}, \ \
\omega(r) \sim  \frac{J}{r^3},  \\
a_{0}(r) &\sim & -\frac{Q}{\pi r^2}, \ \
a_{\varphi}(r) \sim  \frac{{\mu}}{\pi r^2}. \nonumber
\label{expan_infty}
\end{eqnarray}

The constants $J$, $Q$, $\alpha$, $\beta$, and $\mu$ describing the asymptotic behavior are related to the global charges of the black holes.

In $G_5=1$ units, the electric charge and the angular momentum of the black hole are
\begin{equation}
Q 
= - \frac{1}{2} \int_{S_{\infty}^{3}} *F 
\ ,  \ \
J = \int_{S_{\infty}^{3}} *\nabla \eta_{(1)} = \int_{S_{\infty}^{3}} *\nabla \eta_{(2)}.
\label{charges}
\end{equation}

The parameters $\alpha$ and $\beta$ are related to the total mass $M$ of the black hole. We calculate the mass using the Ashtekar--Magnon--Das conformal mass
definition \cite{Ashtekar:1984zz,Ashtekar:1999jx}, obtaining
\begin{eqnarray}
\label{mass-ct}
M = -\frac{\uppi}{8}\frac{(3\alpha+\beta)}{L^2}~.
\end{eqnarray}

Note that this definition of the mass ignores the 
Casimir contribution to the total mass, $M_0=\frac{3\uppi}{32}L^2$, as~defined in \cite{Balasubramanian:1999re}. Since we will always compare configurations with a fixed AdS length, this term will just be a constant shift in the values of the mass.

Finally, from the magnetic part of the gauge field, we can calculate the parameter 
\begin{eqnarray}
\mu_{(k)} = \lim_{r\to\infty} \pi r^2 A_{\varphi_{(k)}} \equiv \mu,
\end{eqnarray}
  which is the magnetic moment of the black hole.

At the event horizon, $r=r_H$, the Ansatz (\ref{metric}) satisfies the Killing horizon conditions, which implies $f(r_H)=0$ in the non-extremal case. A series expansion around the horizon reveals the following behavior in the case of non-degenerate horizons: 
\begin{small}
\begin{eqnarray}
&&f(r) \sim f_2 (r-r_H)^2~, \ \
m(r) \sim m_2 (r-r_H)^2~, \ \
n(r) \sim n_2 (r-r_H)^2~, \  \\
&&\omega(r) \sim \Omega_H r + \omega_2(r-r_H)^2/2, \nonumber \ \
a_{0}(r) - V_{0} \sim a_{0, 2} (r-r_H)^2~, \ \
a_{\varphi}(r) - V_{\varphi} \sim  a_{\varphi, 2} (r-r_H)^2. 
\end{eqnarray}
\end{small}
\indent These constants $f_2$, $m_2$, $n_2$, $\Omega_H$, $V_{0}$, $V_{\varphi}$, $a_{0, 2}$ and $a_{\varphi, 2}$ are not free, but they are related to the mass, angular momentum and electric charge in a non-trivial way. The parameter $\Omega_H$ is the angular velocity of the event horizon.

Extremal black holes with a degenerate horizon satisfy the conditions $r_H=0$ and $\partial_r f(r_H)=0$.
These two conditions result in a different behavior of the functions near the horizon with respect to the non-extremal case. The next-to-leading terms now contain non-integer exponents:
\begin{equation}
\begin{array}{ll}
f(r) \sim f_0 r^{v_0}~, \ \
m(r) \sim m_0 r^{v_1}~, \ \
n(r) \sim n_0 r^{v_2}~, \ \ \omega(r) \sim \Omega_H r, \\
a_{0}(r) - V_{0} \sim a_{(0, 1)}r^{v_3}~, \ \
a_{\varphi}(r) - V_{\varphi} \sim a_{(\varphi, 1)}r^{v_3}.
\end{array}
\end{equation}

The regularity of the extremal solution implies that these exponents satisfy $v_0>4$, $v_1>2$, $v_2>2$ and $v_3>2$ together with $3v_0-2v_1-v_2=6$. These exponents are not free, they depend implicitly on the parameters of the extremal black holes, e.g.,
on $J$, $Q$ and $L$. 

\textls[-15]{To study the horizon properties, we define several quantities. For example, the angular momentum} stored behind the horizon can be calculated using the Komar integral
\begin{eqnarray}
J_{{H}} = \int_{{\cal H}} *\nabla \eta_{(1)}=\int_{{\cal H}} *\nabla \eta_{(2)}=
\lim_{r \to r_{H}} \frac{\uppi}{16} r^3\sqrt{\frac{m n^3}{f^5}}\left[\omega - r\omega'\right]= 
-\frac{\uppi}{16}r_H^5 \sqrt{\frac{m_2 n_2^3}{f_2^5}} \omega_2,
\label{Hang}
\end{eqnarray}
where $\omega_2=\lim_{r \to r_{H}}\omega''$.
We also define the mass of the horizon using the standard Komar integral (note this integral diverges when evaluated at the AdS$_5$ boundary, which forces us to use the previous definition of the total mass in order to get a regular result)
\begin{equation}
\begin{array}{ll}
M_{H} =&- \frac{3}{2} \int_{{\cal H}}*\nabla \xi = 
\lim_{r \to r_{H}} \frac{3\uppi}{16} r^3\sqrt{\frac{m n}{f^3}}
\left[\frac{n\omega}{f}\left(\frac{\omega}{r}-\omega' \right) 
+ f'\left(1+\frac{r^2}{L^2}\right) + \frac{2rf}{L^2}\right]  \\
&= \frac{3\uppi}{8}r_H^2 \sqrt{\frac{m_2 n_2}{f_2}} (1+\frac{r_H^2}{L^2}) + 3 \Omega_H J_H.
\label{Hmass} 
\end{array}
\end{equation}

The size of the horizon can be calculated by measuring its area, which is
\begin{equation}
A_{H}=\int_{{\cal H}} \sqrt{|g^{(3)}|}=
2\uppi^2 r_{H}^{3} \lim_{r \to r_{H}}
\sqrt{\frac{m^{2} n}{f^{3}}} = 2\uppi^2r_H^3\sqrt{\frac{m_2^2 n_2}{f_2^3}}.\label{hor_area}
\end{equation} 

Another quantity useful to characterize degenerate black holes is the surface gravity at the~horizon:
\begin{equation}
\kappa = \sqrt{\frac{1}{2}}|\nabla\zeta|_{\cal H}
=\frac{f_2}{r_H\sqrt{m_2}}(1+\frac{r_H^2}{L^2}).
\label{hor_kappa}
\end{equation} 

Note that this quantity vanishes in the extremal case.
According to black hole thermodynamics, we~define the entropy of the black hole as $S=\frac{1}{4} A_H$, and the surface temperature as $T = \frac{1}{2\uppi}\kappa$. 

Finally, let us note here that combining Equations (\ref{Hmass}) with (\ref{hor_area}) and (\ref{hor_kappa}), one can see that the horizon mass of these black holes satisfy the following relation 
\begin{equation}
\frac{2}{3}M_H = 
TS
 + 2\Omega_H J_H.
\label{horizon_smarr}
\end{equation}

In the extremal case, it implies the proportionality of the horizon mass with respect to the horizon angular momentum: $M_H=3\Omega_H J_H$.

Expression (\ref{horizon_smarr}) resembles the Smarr formula, but instead of the total mass and the total angular momentum, it is written in terms of the horizon mass and the horizon angular momentum, and it does not depend on the electric charge of the configuration.
In the case of the global charges, such~a~simple expression of the Smarr formula is not available: 
apart from the introduction of a term with the electric charge, the AdS asymptotics of these solutions require the introduction of an additional pressure-volume term related with the cosmological constant, as it has been shown in \cite{Kastor:2009wy,Cvetic:2010jb,Dolan:2013ft,Altamirano:2014tva}.

\section{Excited Black Holes} \label{s3}

By combining the black hole Ansatz we described above, with the field Equations (\ref{Einstein_equation}) and (\ref{Maxwell_equation}), the problem of obtaining the black hole solutions can be reduced to solving a system of ordinary differential equations. This minimal system of ordinary differential equations has been presented previously in \cite{Blazquez-Salcedo:2016rkj} and we will not repeat it here. 

The solutions are then generated numerically. They extend from the outer event horizon up to the AdS$_5$ boundary, and satisfy the boundary conditions and series expansions we just presented at those two extremes. The numerical solver we use to integrate the system of differential equations
implements 
a collocation method for boundary-value ordinary differential equations (COLSYS), 
and is equipped with an adaptive mesh selection procedure \cite{COLSYS1,COLSYS2}.
The typical size of the mesh is around $10^3$--$10^4$ points, and the solutions have a relative accuracy of $10^{-8}$--$10^{-10}$,  
which we test by evaluating several constraints from the field equations, conserved charges and relations at the boundaries, such as the relation (\ref{horizon_smarr}). Note this procedure has been successfully used before to study other regimes of the EMCS-AdS theory \cite{Blazquez-Salcedo:2016rkj}.

\subsection{Branch Structure of the $T = 0$ Black Holes for $\lambda>2$ \label{sect3-1}}
 
For the study of the extremal black holes, it can be useful to compare the global results from the numerics with the near-horizon formalism, which allows for obtaining semianalytical results for the properties of the horizon \cite{Sen:2005wa,Astefanesei:2006dd,Goldstein:2007km,Kunduri:2007qy}. 
These solutions have been studied previously in \cite{Blazquez-Salcedo:2016rkj,Blazquez-Salcedo:2013muz,Blazquez-Salcedo:2015kja}, and we shall briefly mention here how to construct them. The near-horizon region of the cohomogeneity-1 black holes we are considering present an enhanced isometry group with respect to the full global solutions, essentially given by the product of AdS$_2$ with a squashed sphere $S^3$ \cite{Kunduri:2007qy}. Hence, the metric and the gauge field can be parametrized in a simple way in terms of some indeterminate constants:
\begin{eqnarray}
 \label{NHmetric}
&&ds^2 = v_1(\frac{dr^2}{r^2}-r^2dt^2) +v_2 \left( d\theta^2 +(1-\eta)\sin^2\theta \cos^2\theta \left( d \varphi_1  - d \varphi_2\right)^2 \right)  \nonumber \\
&&
+v_2\eta \left( \sin^2\theta \left( d \varphi_1 -\alpha r dt \right)^2
+               \cos^2\theta \left( d \varphi_2 -\alpha r dt \right)^2 \right),  \\
&&A = -q dt + p \left( \sin^2\theta d \varphi_1 + \cos^2\theta d \varphi_2 \right). \nonumber
\end{eqnarray}

With this Ansatz, it is possible to reduce the field Equations (\ref{Einstein_equation}) and (\ref{Maxwell_equation}) to a set of algebraic relations for these constants, and obtain a relation between the area, the angular momentum and the electric charge. For the specific expression of these relations, we refer the reader to \cite{Blazquez-Salcedo:2016rkj}.

Let us now continue by presenting the numerical results for extremal black holes. In the following, we will focus the analysis on the case $\lambda=5$. We will also fix the electric charge of these black holes to $Q=-2.72$ and the AdS$_5$ length to $L=10$. However, similar features are found for other arbitrary values of these constants, provided that $\lambda>2$. 

\begin{figure}[H]
     \centering
\begin{subfigure}[b]{0.45\textwidth}
     \centering
\includegraphics[width=52mm,scale=0.5,angle=-90]{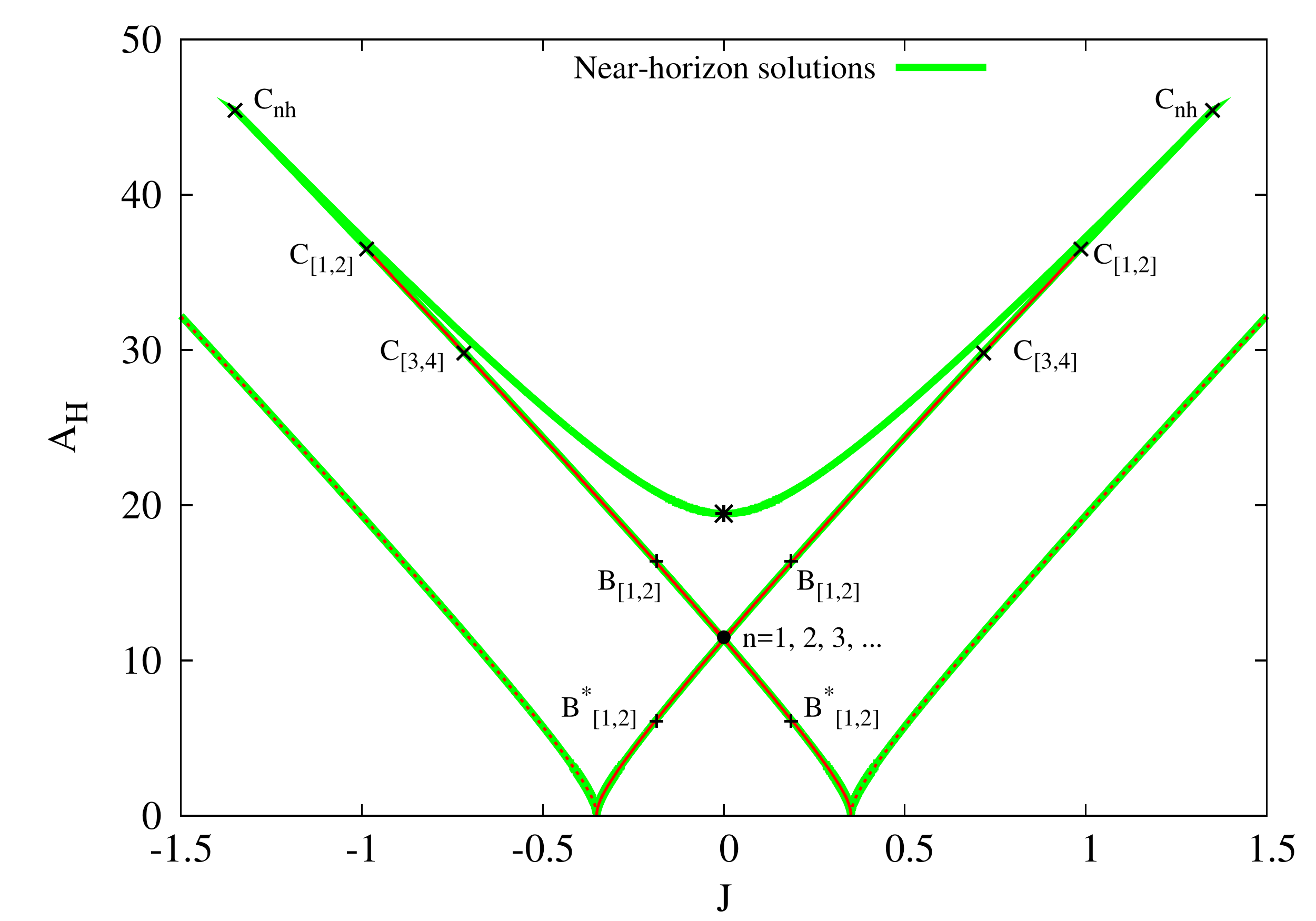}
         \caption{}
\label{plot_AhvsJ}
\end{subfigure}
\begin{subfigure}[b]{0.45\textwidth}
     \centering
\includegraphics[width=52mm,scale=0.5,angle=-90]{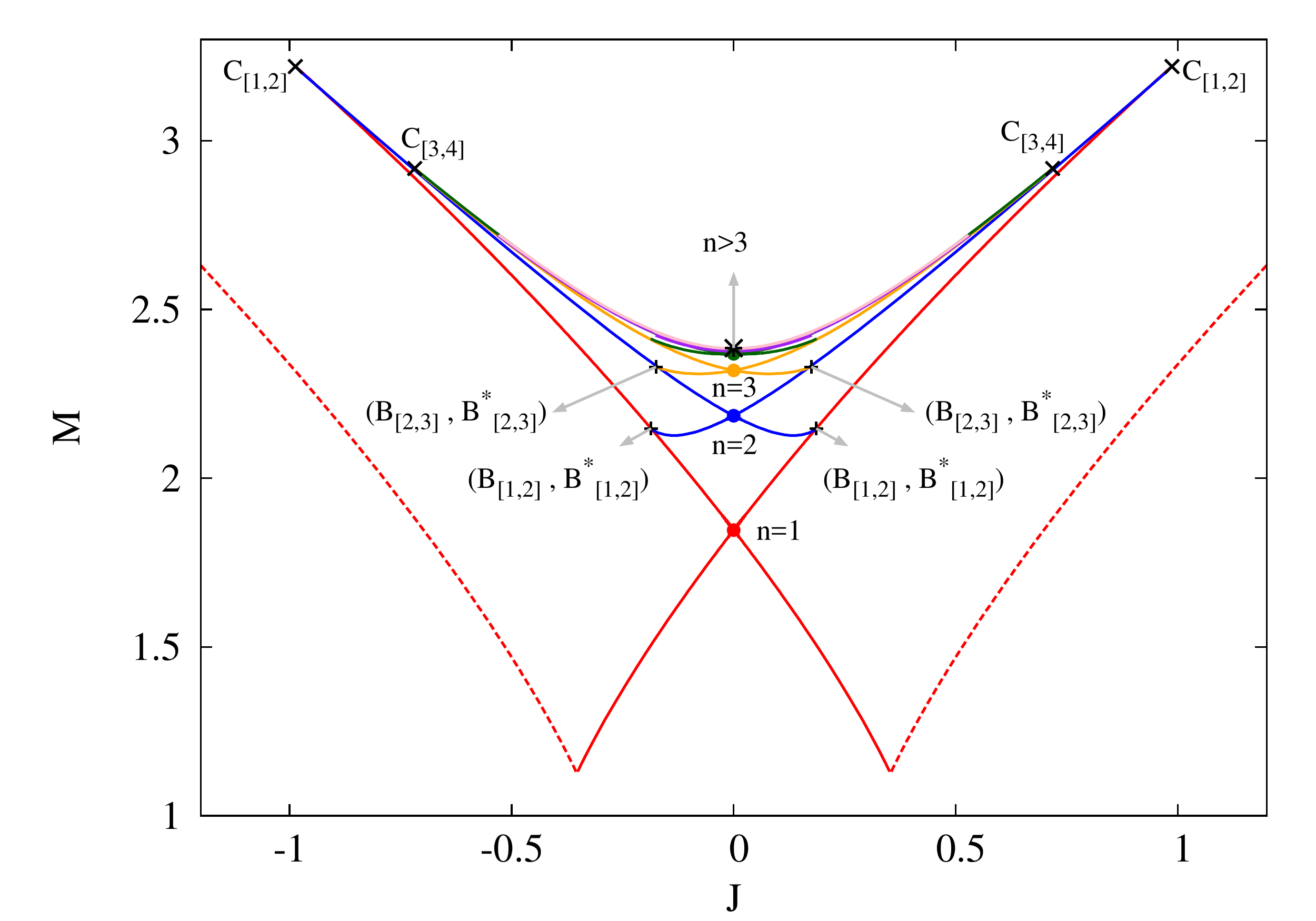}
         \caption{}
\label{plot_MvsJ}
\end{subfigure}
\vspace{-12pt}
\caption{(\textbf{a}) area of the event horizon $A_H$ versus the total angular momentum $J$ for black holes with $\lambda=5$, $Q=-2.72$ and $L=10$. The thick line represents the prediction from the near horizon formalism (green). The thin line represents the global black hole solutions extended to the AdS$_5$ boundary (red). (\textbf{b}) mass $M$ versus the total angular momentum $J$ for the global black holes with the same set of charges.  
In both figures, the dashed red line is the branch of black holes connecting with the extremal MP-AdS solution. The cusps are marked by an $\times$ and denoted by
$C_{[...]}$, the bifurcation points are marked by $+$ and denoted
by $B_{[...]}$. The black star represents the extremal {RN}-AdS solution. The non-static $J=0$ solutions are marked with a black dot in (\textbf{a}), and possess the same value of the area. In (\textbf{b}), they are marked with a dot of the same color of the branch they belong to. 
} \label{f1}
\end{figure}

In Figure \ref{f1}a, we present the area of the event horizon $A_H$ versus the total angular momentum $J$. In this figure, we have included for reference the set of near-horizon solutions as a thick green line. 
The set of global black hole solutions is depicted as a red line, and it is easy to see that these two sets do not always match.  

Since we are exploring black holes with fixed $Q$, the extremal MP-AdS solution would be recovered at large values of the angular momentum $J$, for which these configurations are negligibly charged. Decreasing the value of the angular momentum generates a full branch of solutions we call the extremal Myers--Perry (MP) branch, since it connects  asymptotically with the extremal MP solution. This set extends from $|J|=\infty$ down to some critical value of $|J_c|$ (in Figure \ref{f1}a, this is $|J_c|=0.38$).~At this critical point, the black hole has zero area and the geometry of the horizon changes from AdS$_2 \times S^3$ to AdS$_3 \times S^2$. This type of zero-area configuration is common in the EMCS theories, and, in particular, it~is present in the CPL black hole \cite{Blazquez-Salcedo:2016rkj}.

The extremal MP branch is shown in Figure \ref{f1}a as a dashed red line. Note that, along this branch, the area increases almost linearly with $|J|$. In addition, it is important to note that the near-horizon solutions have a one-to-one correspondence with the global configurations along this branch.

For lower values of $J<|J_c|,$ the situation is more complicated.~The near horizon formalism predicts that the extremal RN solution (marked with a black star) can be connected with a non-static $J=0$ solution of a smaller area. Both solutions are connected through a cusp $C_{nh}$ solution, which possesses a local maximum of the area and $|J|$. 

However, global black hole solutions exhibit a different behavior. This branch is displayed in Figure \ref{f1}a as a solid red line. The angular momentum can be decreased to zero, but, at this point, we~reach a non-static $J=0$ configuration (marked with a black dot, corresponding to $n=1$). The~new branch however doesn't stop here. Instead, the angular momentum can be increased again up to a~maximum value of $|J|$ and $A_H$ at a configuration, we denote as cusp $C_{[1,2]}$ for reasons we will discuss in the following. 

Just by inspecting Figure \ref{f1}a, we could be tempted to think that those are all the global black holes present in the theory. However, this not true. At the cusp $C_{[1,2]},$ another branch of black holes emerges, for which the values of $|J|$ can be decreased again to zero, reaching a different $J=0$ non-static solution ($n=2$). The branch eventually disappears at a branching point $B^*_{[1,2]}$. As a matter of fact, this is just the tip of the iceberg, since this complicated branch structure repeats an infinite number of times, a~feature that can only be appreciated in the global quantities.    

Before continuing with the description of the global quantities, let us note that, from Figure~\ref{f1}a, we~can learn two very interesting facts. First, similar to what was found in 
\cite{Blazquez-Salcedo:2013muz,Blazquez-Salcedo:2015kja}, for the asymptotically flat case, not all near-horizon solutions correspond to global black holes. For example, note that the cusp solution $C_{nh}$ and the branch of near-horizon solutions connecting with the RN-AdS black hole has no global counterpart. Second, and, as a result of the previous observation, the extremal Reissner--Nordstr\"om-AdS black hole is in fact isolated from the rotating and charged extremal black holes, in the sense that it cannot be reached by just decreasing the angular momentum of a rotating and charged extremal configuration. However, it can be connected with rotating and charged configurations only if one allows for $T\neq 0$ black holes (i.e., relaxing the extremality condition). We will come back to this point in the next sections.

In Figure \ref{f1}b, we present the mass $M$ versus the angular momentum.~Note that we are not including the Casimir term into the mass. Again, we can see as a dashed red line the extremal MP branch, which~extends from large $|J|$ down to the critical value of $|J_c|=0.38$, where the mass exhibits a local minimum (but does not vanish).

At the other side of the critical value, we find the infinite branch structure, which, in Figure \ref{f1}b is displayed as solid lines of different colors. Each color matches a different $n$ number. The $n=1$ branch is in red and starts at $|J_c|=0.38$. As the angular momentum decreases along this branch, we reach the $n=1$ non-static $J=0$ black hole (red dot). The branch continues with increasing $|J|$ up to the cusp $C_{[1,2]}$. At the cusp $C_{[1,2]}$, the $n=2$ branch appears, and it includes the $n=2$ non-static $J=0$ black hole (blue dot). Note the mass of this $n=2$ branch is larger than the mass of the $n=1$ branch. The branch disappears at the branching point $(B_{[1,2]},B^*_{[1,2]})$.

On this branching point, there are two different configurations: 
the black hole $B_{[1,2]}$, which belongs to branch $n=1$ (red line), and the black hole $B^*_{[1,2]}$, which belongs to branch $n=2$ (blue line). These~two black holes have the same set of charges $Q$, $J$ and $M$. However, if we go back to Figure \ref{f1}a, we~can see that in fact they have very different horizon properties: the black holes $B_{[1,2]}$ have a larger area than the black holes $B^*_{[1,2]}$. In conclusion, we have non-uniqueness between global charges for these extremal black holes: these two black holes have the same mass, but they possess different near horizon geometry and can be distinguished, for example, by their different area.

In Figure \ref{f1}b, we can appreciate that, along branch $n=2$, another branching point $(B_{[2,3]},B^*_{[2,3]})$ appears. Here, the $n=3$ branch (orange line) is generated. This branch has similar features to the previous branches: it has a non-static $J=0$ solution ($n=3$, orange dot), and goes up to the cusp $C_{[3,4]}$. At this cusp, we find the $n=4$ branch (green line), which also has a non-static $J=0$ solution and new branching points. The structure repeats an infinite number of times. The mass of the branch structure becomes closer and closer to the static mass (marked in Figure \ref{f1}b with a black star).

\begin{figure}[H]
\begin{subfigure}[b]{0.45\textwidth}
\centering
\includegraphics[width=52mm,scale=0.5,angle=-90]{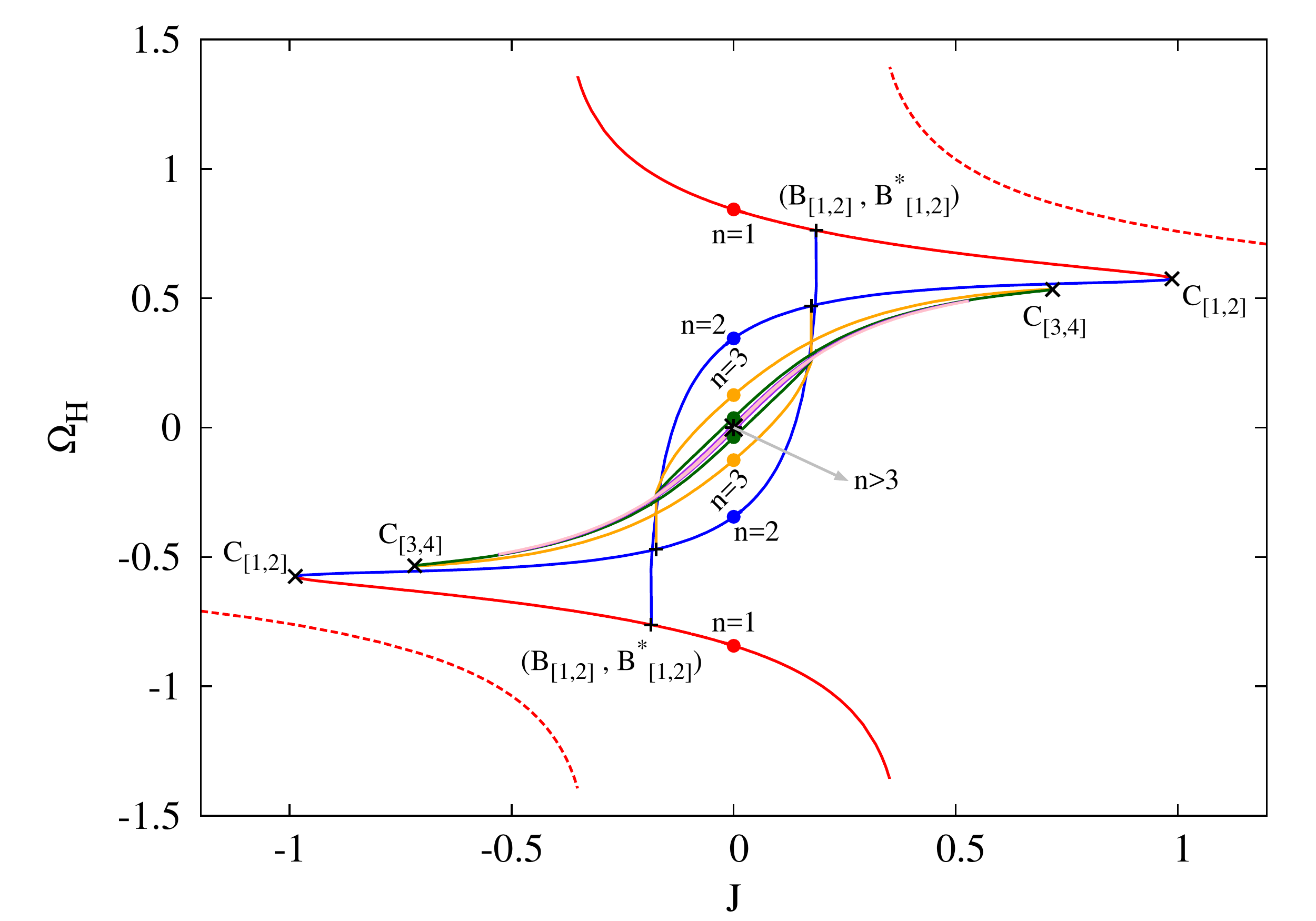}
         \caption{}

\end{subfigure}
\begin{subfigure}[b]{0.45\textwidth}
\centering
\includegraphics[width=52mm,scale=0.5,angle=-90]{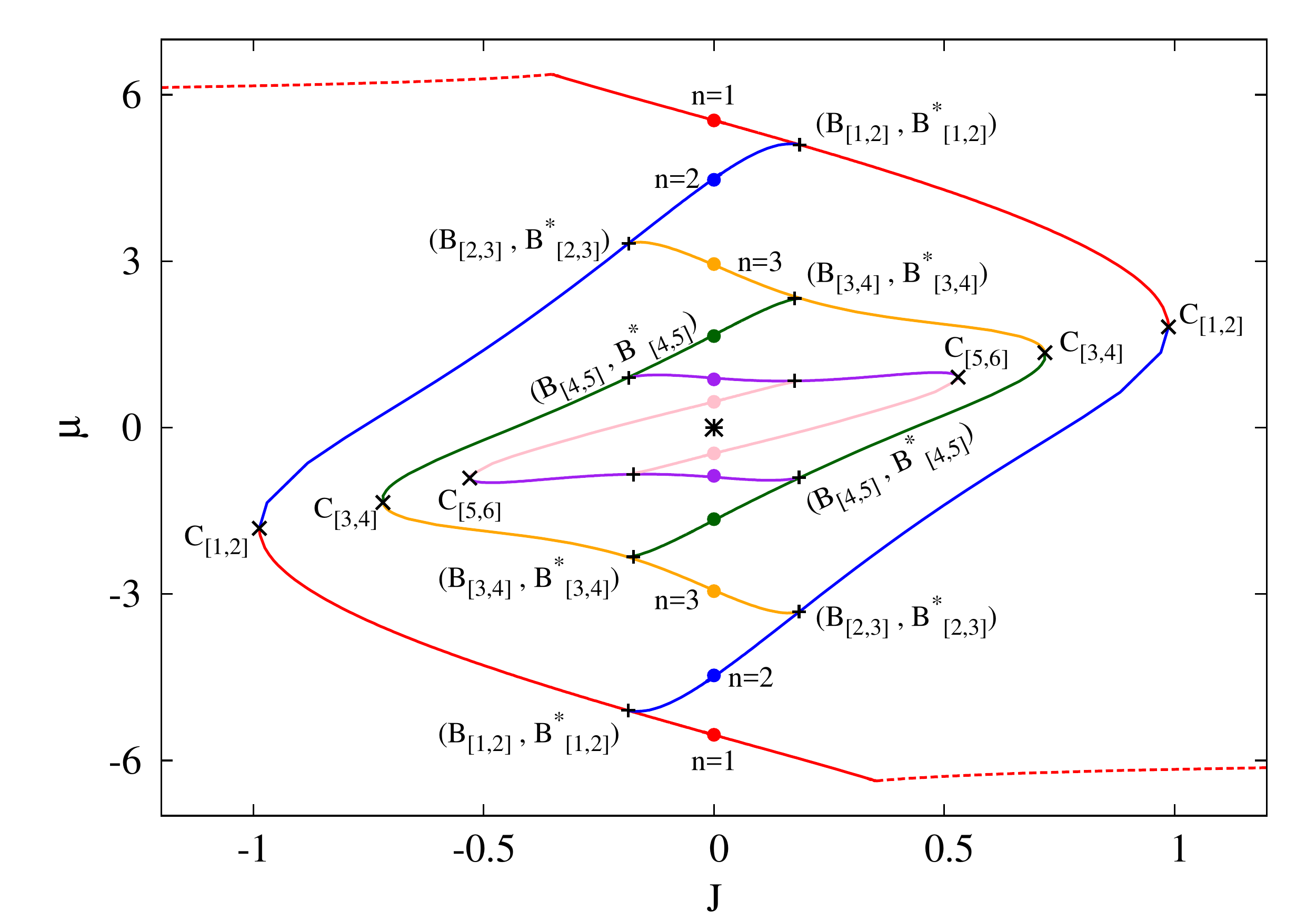}
         \caption{}

\end{subfigure}
\vspace{-12pt}

\caption{(\textbf{a}) horizon angular velocity $\Omega_H$ versus the total angular momentum $J$ for black holes with $\lambda=5$, $Q=-2.72$ and $L=10$; (\textbf{b}) magnetic moment $\mu$ versus the total angular momentum $J$ for the same set of black holes. The dashed red line is the branch of black holes connecting with the extremal MP-AdS solution. The solid lines represent the different branches: in red, the branch containing $n=1$, in blue $n=2$, in orange $n=3$, etc...
The cusps are marked by an $\times$ and denoted by $C_{[...]}$, the bifurcation points are marked by $+$ and denoted
by $B_{[...]}$. The black star represents the extremal RN-AdS solution, and the color dots the non-static $J=0$ solutions. For clarity in (\textbf{b}), we have included only the first six~branches.} \label{f2}
\end{figure}

The mass is not the only property that can be used to distinguish between different branches. In~Figure \ref{f2}a, we present the horizon angular velocity $\Omega_H$ versus the angular momentum $J$. We follow the same notation as in Figure \ref{f1}b.

Interestingly, in Figure \ref{f2}a, we can see that, at the critical solution $|J_c|$ of the vanishing area, the~angular velocity jumps from $\Omega_H$ to $-\Omega_H$. We can also see how the non-static $J=0$ solutions are, in~fact, not~static, since they have a non-zero angular velocity. Similarly, every branch contains non-static $\Omega_H=0$ solutions, with non-zero angular momentum.  

Finally, in Figure \ref{f2}b, we present the magnetic moment $\mu$ versus the angular momentum. In this figure, it is easier to see the branching structure, and, for clarity, we have included only the first six branches: red for $n=1$, blue for $n=2$, orange for $n=3$, green for $n=4$, purple for $n=5$ and pink for $n=6$. At the center of the figure, we find the static black hole (black star) with null magnetic moment. The branch structure is very reminiscent of a demagnetization loop for a ferromagnetic material, with the magnetic moment and the angular momentum playing the roles of the induced magnetic flux and the magnetizing force, respectively. Ferromagnetic-like behavior of black holes are known in other contexts, like in dyonic black holes \cite{Cai:2014oca}. 

From Figure \ref{f2}b, we can infer that the branch structure follows these rules:

\begin{itemize}[leftmargin=*,labelsep=5mm]
\item	The $n=2j+1$ branch (with $j \in \mathbb{Z}$) connects with the $2j+2$ branch via a single solution, the cusp $C_{[2j+1,2j+2]}$.
\item	Every branch of number $n$ contains a branching point $(B_{[n,n+1]},B^*_{[n,n+1]})$ at some $J\neq 0$. At these points, there is non-uniqueness with respect to the global quantities ($Q$, $M$, $J$) of extremal black holes, but the solutions can be distinguished by their horizon properties. This means it is possible to jump from branch $n$ to $n+1$ through a branching point keeping all global quantities continuous, but this requires a jump of the entropy $\Delta S <0$.
\item	Each branch of number $n \geq 1 $ contains one non-static $J=0$ solution, which can be labeled by the number $n$ of the branch.
\item	Similarly, each branch of number $n>1$ also contains one non-static $\Omega_H=0$ solution, which can also be labeled with the $n$ number.
\end{itemize}

\begin{figure}[H]
         \begin{subfigure}[b]{0.45\textwidth}
              \centering
     \includegraphics[width=52mm,scale=0.5,angle=-90]{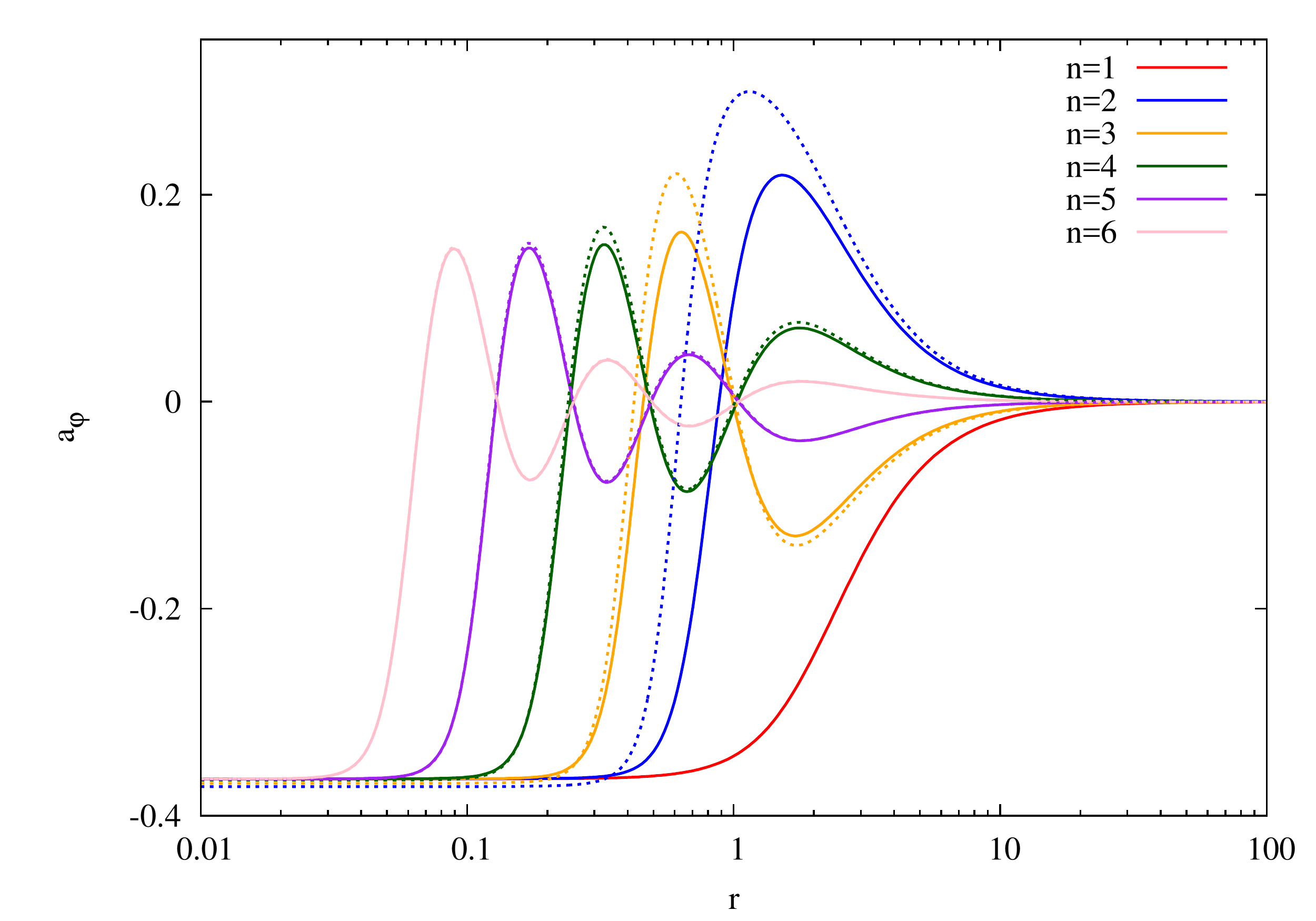}
         \caption{}       
     \end{subfigure}
     \begin{subfigure}[b]{0.45\textwidth}
          \centering
     \includegraphics[width=52mm,scale=0.5,angle=-90]{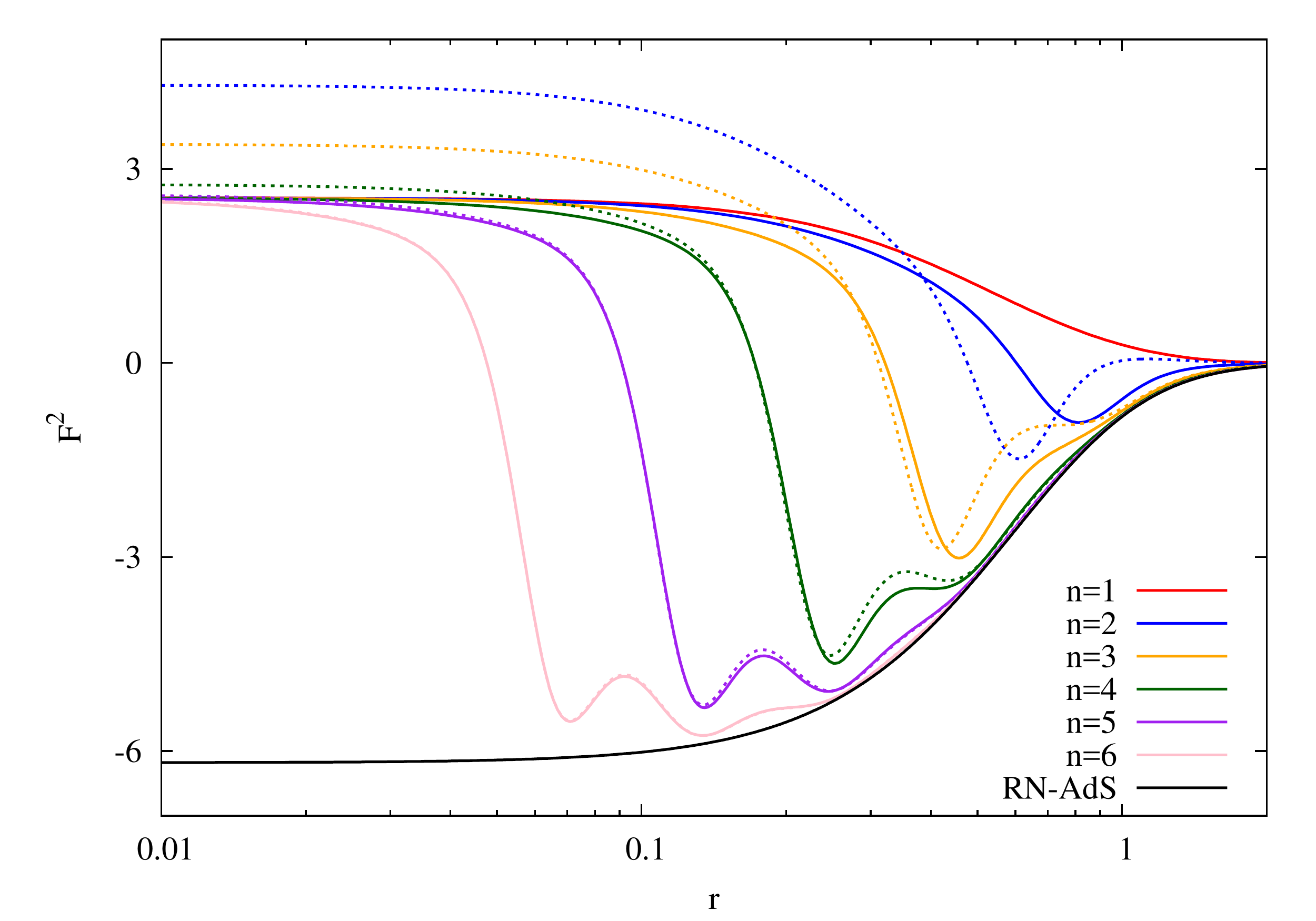}
         \caption{}
      
     \end{subfigure}
\vspace{-12pt}
     \caption{(\textbf{a}) the function $a_{\varphi}$ versus the radial coordinate $r$ for different excited black holes. Here, $\lambda=5$, $L=10$, $Q=-2.72$. The solid lines represent non-static $J=0$ solutions, and the dashed ones non-static $\Omega_H=0$ solutions. In (\textbf{b}), we show a similar plot for the invariant $F^2$ and the same set of configurations.}

         \label{f3}

\end{figure}

The integer numbers of the non-static $J=0$ and $\Omega_H=0$ configurations possess a physical meaning. In Figure \ref{f3}a, we display the profiles for the function $a_{\varphi}$, and, in Figure \ref{f3}b, the invariant $F^2$. With~solid lines, we show the non-static $J=0$ black holes, and with dashed lines the non-static $\Omega_H=0$ black holes for $n=1, ..., 6$. In these figures, we can see how the integer number $n$ is related with the structure of the radial profile in the Ansatz functions. We observe that the $n$ number counts the number of nodes of the magnetic part of the gauge field $a_{\varphi}$ as function of the quasi-isotropic radial coordinate~$r$. The~same holds for the inertial dragging $\omega$ (since the CS interaction is essentially coupling the magnetic part of the gauge field with the inertial dragging).

%%%%%%%%%%%%%%%%%%%%%%%%%%%%%%%%%%%%%%%%%%%%%%%
%%%%%%%%%%%%%%%%%%%%%%%%%%%%%%%%%%%%%%%
This node structure manifests itself at the level of electro-magnetic invariants, like $F^2$ shown in Figure \ref{f3}b. Hence, the set of $J=0$ configurations can be understood as black holes with a radially excited magnetic field. The mass and magnetic moment of these solutions do not vanish, they are discrete functions of the excitation number $n$. The lowest excitation, $n=1$ with only one node in the radial profile, possesses the lowest mass (energy) and the largest magnetic moment. Increasing the excitation number increases the number of oscillations, which increases the mass of the black hole and decreases the magnetic moment. As $n \to \infty$, the oscillatory behavior of the magnetic part becomes damped, and the extremal RN-AdS solution is recovered.

As a final note in this section, we would like to point out that the properties of these configurations we have described here are very similar to the ones found in \cite{Blazquez-Salcedo:2013muz,Blazquez-Salcedo:2015kja} for asymptotically flat black holes. The branch structure of the extremal black holes in the large CS coupling is qualitatively unchanged by the introduction of the cosmological constant and the AdS asymptotics. Hence, radially excited solutions exist also with the AdS asymptotics.
This makes these excited configurations potentially more interesting in the context of the AdS/CFT correspondence. In this sense, it is worth mentioning that these radially excited configurations (found for fixed values of $Q$ and $J=0$), differ on the stress-energy tensor at the AdS boundary (since they possess different values for example of the total mass), but~all of them possess the same value of the area (entropy), with a fixed near-horizon geometry that depends only on $Q$. 
For instance, in Figure \ref{f3}, we can see how the near-horizon region of the non-static $J=0$ black holes is pushed away from the AdS boundary when the excitation number is increased. On~the other hand, we have seen that the opposite situation is also possible: at the branching points, we can find pairs of solutions with the same properties at the AdS boundary (same electric charge, angular~momentum, total mass...). However, these two solutions have different values of the entropy, given~that each of them possesses a different near-horizon geometry.

Thus far, we have described the properties of extremal black holes. In the next section, we will focus on these sets of excited non-static black holes, and we will study the effect of the temperature on~them.

\subsection{Thermodynamic Properties of the Non-Extremal Excited Black Holes}

\subsubsection{Non-Static $J=0$ Black Holes}

In this section, we will discuss some thermodynamic properties of the non-static $J=0$ black holes and their relation with the static black hole. In particular, we are interested in the thermodynamic stability of these excited configurations. Since their total angular momentum vanishes, it seems natural to ask what is the difference with respect to the Reissner--Nordstr\"om-AdS black hole.

A canonical ensemble is characterized by the temperature of the black holes $T$, electric charge $Q$, and angular momentum $J$.~In this ensemble, the stability with respect to variations of the thermodynamic variables of the black hole can be performed by studying the entropy as a function of the charges. A criterion for stability is that the entropy has to be a concave function of the corresponding extensive thermodynamical parameters. In the case of the canonical ensemble, the local stability condition implies that the specific heat has to be positive \cite{Caldarelli:1999xj,Dolan:2013yca}. The specific heat of a canonical ensemble can be calculated by taking variations of the entropy with respect to the temperature for constant $Q$ and $J$:
\begin{equation}
C_p = T \frac{\partial S}{\partial T}|_{J,Q}.
\end{equation}

However, to ensure global thermodynamic stability, the positivity of the specific heat is not enough, and one has to require that the moment of inertia
\begin{equation}
\varepsilon = \frac{\partial J}{\partial \Omega_H}|_{T,Q}~
\end{equation}
is not negative. 
Note that, in $D>4$, the moment of inertia is a tensorial quantity: $\varepsilon_{ij} = \frac{\partial J_i}{\partial \Omega_{H j}}|_{T,Q}$, and~the stability condition translates into the positivity of the eigenvalues of this tensor. However,~in our particular case, both angular momenta are equal in magnitude and we can just consider the quantity $\varepsilon = \frac{\partial J}{\partial \Omega_H}|_{T,Q}$.
In the following discussion, we will focus on the local stability of canonical ensembles, but we will comment on the issue of the global stability at the end of this section. 

The stability of the Reissner--Nordstr\"om-AdS black hole has been well studied in the literature~\cite{Mitra:1999ge}. In Einstein--Maxwell--Chern--Simons theory, the Reissner--Nordstr\"om-AdS black hole is always a possible static solution of the system, which is described by two free parameters, the temperature $T$ and the electric charge $Q$. It is well known that, when considering a set of solutions with constant electric charge different phase transitions are found, with regions where the thermodynamic stability changes (i.e., the sign of the specific heat $C_p$ changes) \cite{Mitra:1999ge,Caldarelli:1999xj,Dolan:2013yca}. 

Essentially, a canonical ensemble of Reissner--Nordstr\"om-AdS black holes, with a fixed value of $Q$, possesses three branches. The first branch is composed of small black holes, for which $C_p>0$ and the area and mass are typically small. This branch ends at some finite temperature $T_a$ where $C_p$ changes sign, having at this temperature a critical solution that delimits stable and unstable configurations. The second branch is composed of thermodynamically unstable black holes and exists between $T_a$ and $T_b<T_a$, with larger values of the area than the small black hole branch. The third branch, with thermodynamically stable black holes, is found for $T>T_b$ and large values of the area and the mass. This is the branch of large stable black holes.

\begin{figure}[H]
    \begin{subfigure}[b]{0.45\textwidth}
          \centering
     \includegraphics[width=52mm,scale=0.5,angle=-90]{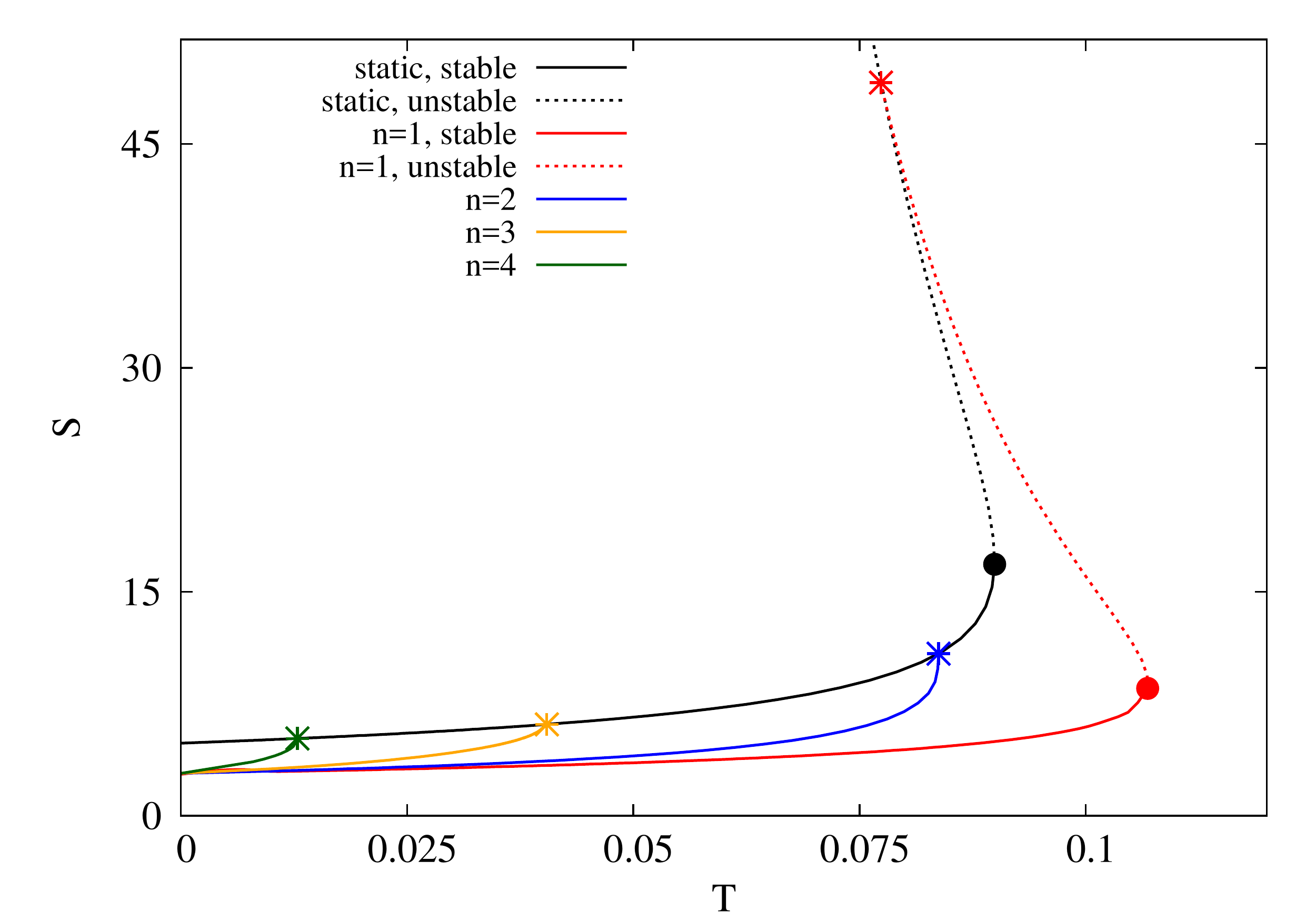}
         \caption{$\lambda=5$}
    \end{subfigure}
     \begin{subfigure}[b]{0.45\textwidth}
          \centering
     \includegraphics[width=52mm,scale=0.5,angle=-90]{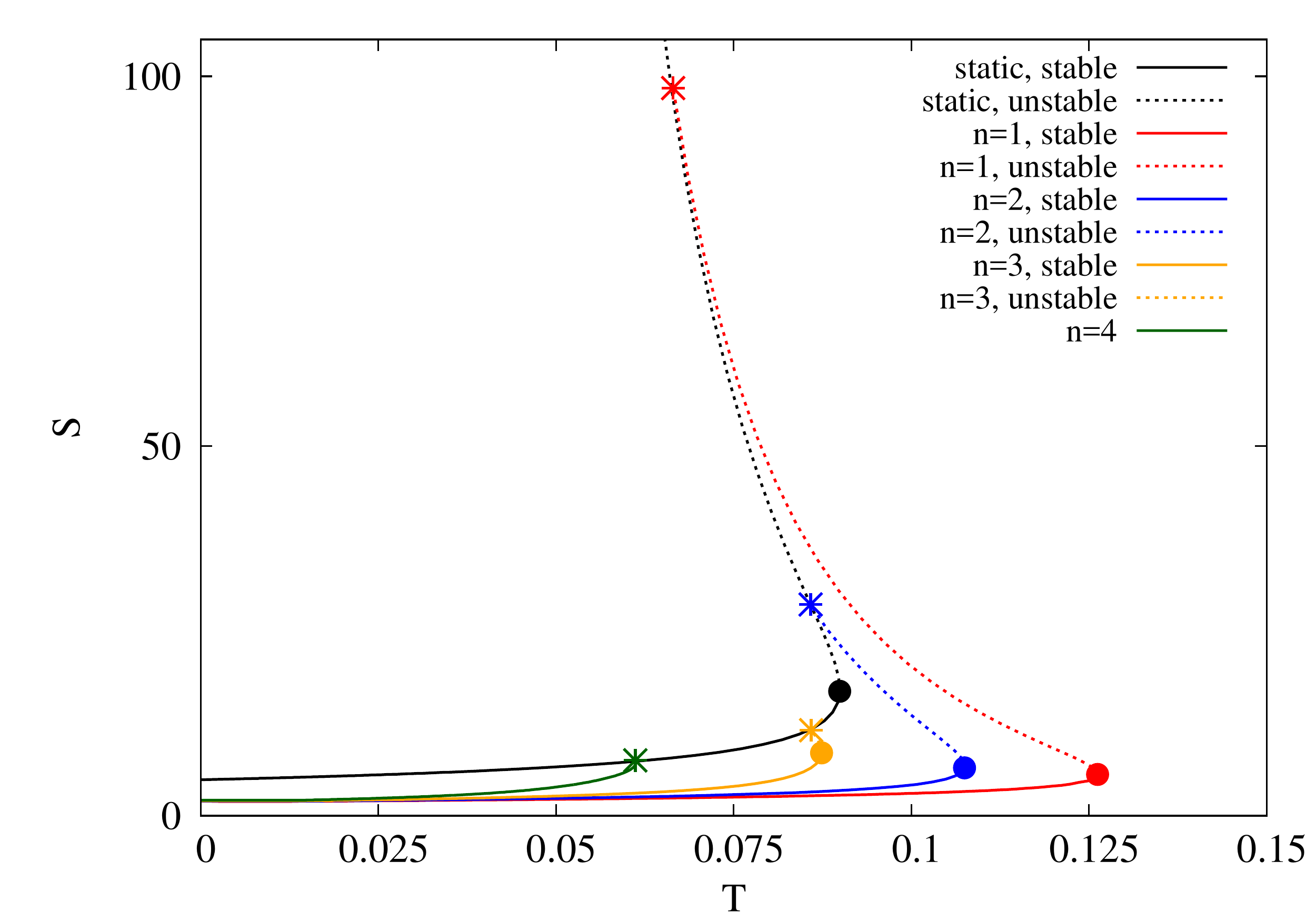}
         \caption{$\lambda=10$}
             \end{subfigure}
\vspace{-12pt}
     \caption{Entropy $S$ versus temperature $T$ for $J=0$ black holes with $Q=-2.72$ and $L=10$. In~(\textbf{a}), we show the $\lambda=5$ case, and in (\textbf{b}) $\lambda=10$. In both figures, the black solid line shows the Reissner--Nordstr\"om-AdS black holes with $C_p >0$, while the black dashed line shows the ones with $C_p <0$. For the non-static $J=0$ black holes, each color represents a canonical ensemble with different excitation number $n$: $n=1$ in red, $n=2$ in blue, $n=3$ in orange, and $n=4$ in green. Solid lines represent thermodynamically stable branches, and dashed lines the unstable ones. For simplicity, we~only include the first four excitations. Dots represent the critical solution where the branch changes stability. The stars mark the points where the non-static $J=0$ black holes merge with the static solution.}

        \label{f4}

\end{figure}

In Figure \ref{f4}, we show the entropy versus the temperature for the canonical ensemble of Reissner--Nordstr\"om-AdS black holes with $Q=-2.72$ and $L=10$. In the figure, we can see the branch of small black holes (solid black line) with thermodynamic stability. This branch ends at $T_a=0.09$ (black dot). At this point, the branch of unstable black holes emerges with increasing entropy as the temperature decreases (dashed black line). The branch ends at $T_b=0.047$, and the third branch of large black holes appears for larger values of the area, which, for simplicity, is not shown in this figure. Note that the Reissner--Nordstr\"om-AdS set is the same in Figure \ref{f4}a for $\lambda=5$ and Figure \ref{f4}b for $\lambda=10$, since the static solutions are not affected by the CS term. 

%%%%%%%%%%%%%%%%%%%%%%%%%%%%%%%%%%%%%%%%%%%%%%%
%%%%%%%%%%%%%%%%%%%%%%%%%%%%%%%%%%%%%%%

The non-static excited solutions are found for a fixed value of Q, and all of them have vanishing angular momentum. It is possible to increase the temperature of these solutions, moving them away from extremality. The resulting set of solutions will form a canonical ensemble, characterized by $J=0$, a fixed value of $Q$, and fixed excitation number $n$.

We can ask ourselves how is the stability of these excited canonical ensembles. Consider the first excitation ($n=1$) of non-static $J=0$ black holes.  In Figure \ref{f4}a, we show such an excited canonical ensemble ($\lambda=5$, $Q=-2.72$, $L=10$, $J=0$, $n=1$) as a solid red line. Starting at the extremal configuration in $T=0$, we see that, as the temperature rises, the entropy of the ensemble increases, meaning that this part of the branch is stable. In particular, the extremal $n=1$ excited black hole is thermodynamically stable. This branch ends at a critical configuration with temperature $T=0.107$ (red dot) where the stability changes. Here, a secondary branch of unstable excited black holes emerges (dashed red line). This behavior is similar to what happens in the static ensemble around $T_a$. However, the unstable $n=1$ branch does not exist for arbitrary values of $S$, eventually merging with the unstable static branch (red star) around $T=0.0774$. This is the end point of the non-static $J=0$, $n=1$ canonical ensemble. Note the merging of the $n=1$ ensemble and the static ensemble happens at a~thermodynamically unstable configuration.

Nevertheless, higher excitations can exhibit a somewhat different behavior. Consider in Figure~\ref{f4}a, at $T=0$, the non-static $J=0$ extremal black hole with $n=2$. If we increase the temperature, we generate another canonical ensemble (blue line). This ensemble always has entropy increasing with the temperature; therefore, this set is thermodynamically stable. Eventually, the branch merges with the static ensemble, which happens on the branch of static small black holes, which is also thermodynamically stable. \enlargethispage{0.5cm}

%%%%%%%%%%%%%%%%%%%%%%%%%%%%%%%%%%%%%%%%%%%%%%%

%%%%%%%%%%%%%%%%%%%%%%%%%%%%%%%%%%%%%%%

In Figure \ref{f4}a, we also include $n=3$ (orange) and $n=4$ (green), both showing the same features as $n=2$. For these specific values of the Chern--Simons coupling and the electric charge, the canonical ensembles of non-static $J=0$ black holes with $n>1$ are always thermodynamically stable. As their temperature is increased, they eventually merge with the branch of static small black holes (stable). For this value of the Chern--Simons coupling, the canonical ensemble with $n=1$ is special, in the sense that it has a subset of unstable black holes, very much like the static ensemble.

%%%%%%%%%%%%%%%%%%%%%%%%%%%%%%%%%%%%%%%%%%%%%%%

%%%%%%%%%%%%%%%%%%%%%%%%%%%%%%%%%%%%%%%

However, these features are not generic. The stability of each excited canonical ensemble depends on the value of the Chern--Simons coupling and the electric charge. For instance, consider Figure~\ref{f4}b, where we show such excited canonical ensembles for the same charge as in Figure \ref{f4}a, but now for $\lambda=10$. Note that the $n=1$ still presents a subset of unstable black holes, but now $n=2$ and also a small portion of the $n=3$ ensembles possess a critical solution with maximum temperature, and~unstable branches that merge with the static branch. Even more, the end points of the excited unstable branches depend on the Chern--Simons coupling.

For a fixed value of the electric charge, any excited canonical ensemble of higher $n$ can contain an unstable branch provided that the Chern--Simons coupling is large enough, and, on the contrary, for values of the Chern--Simons coupling closer to $\lambda=2$, all the excited canonical ensembles possess thermodynamic stability.

Another interesting feature we have observed in our results is that all the extremal excited black holes obtained in our work present local thermodynamic stability, independently of the CS coupling, electric charge, and excitation number. Hence, we conjecture that this is a generic feature of these non-static $J=0$ solutions. 

\begin{figure}[H]
         \centering
     \begin{subfigure}[b]{0.45\textwidth}
          \centering
     \includegraphics[width=50mm,scale=0.5,angle=-90]{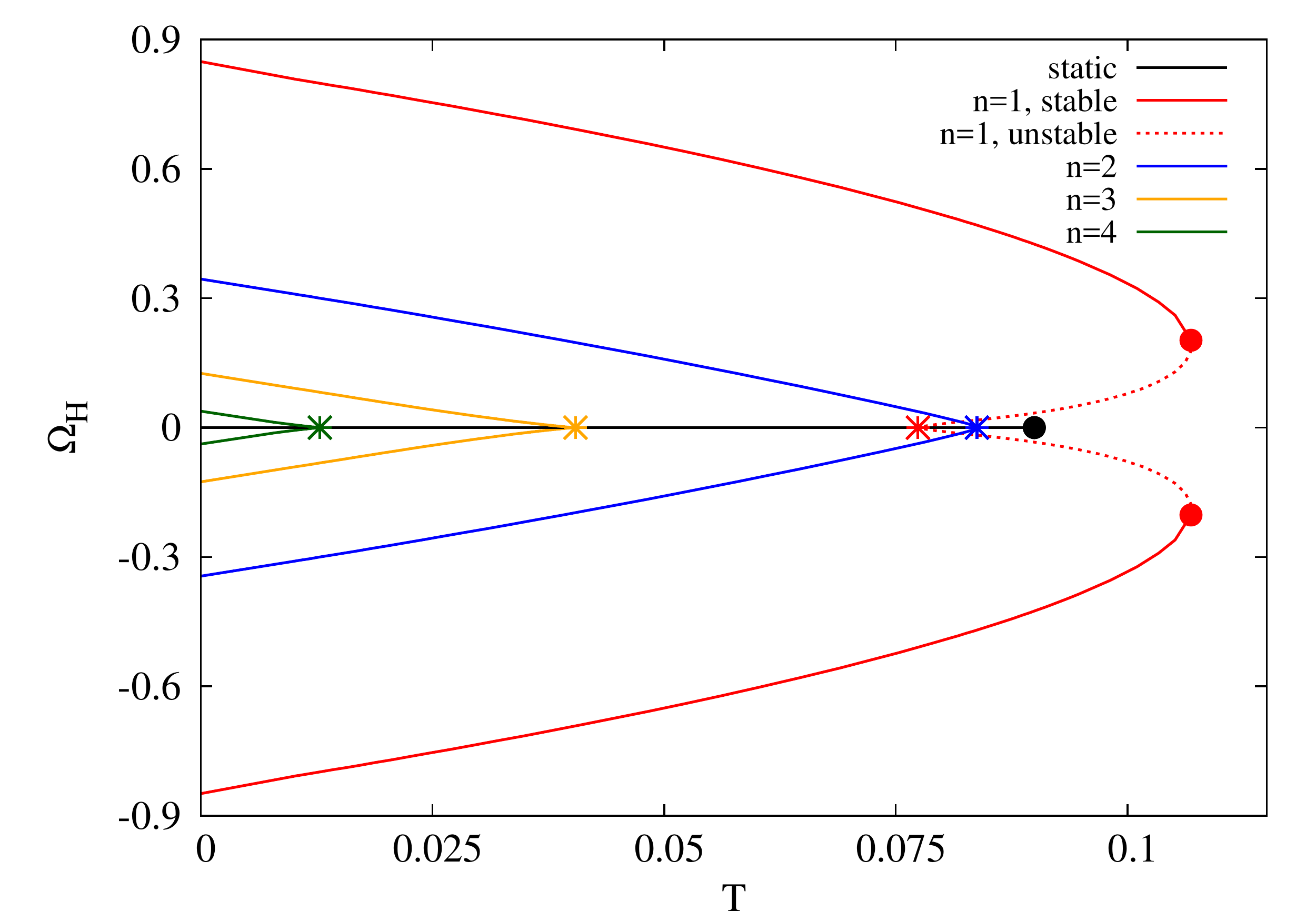}
         \caption{}
         \end{subfigure}
     \begin{subfigure}[b]{0.45\textwidth}
          \centering
     \includegraphics[width=50mm,scale=0.5,angle=-90]{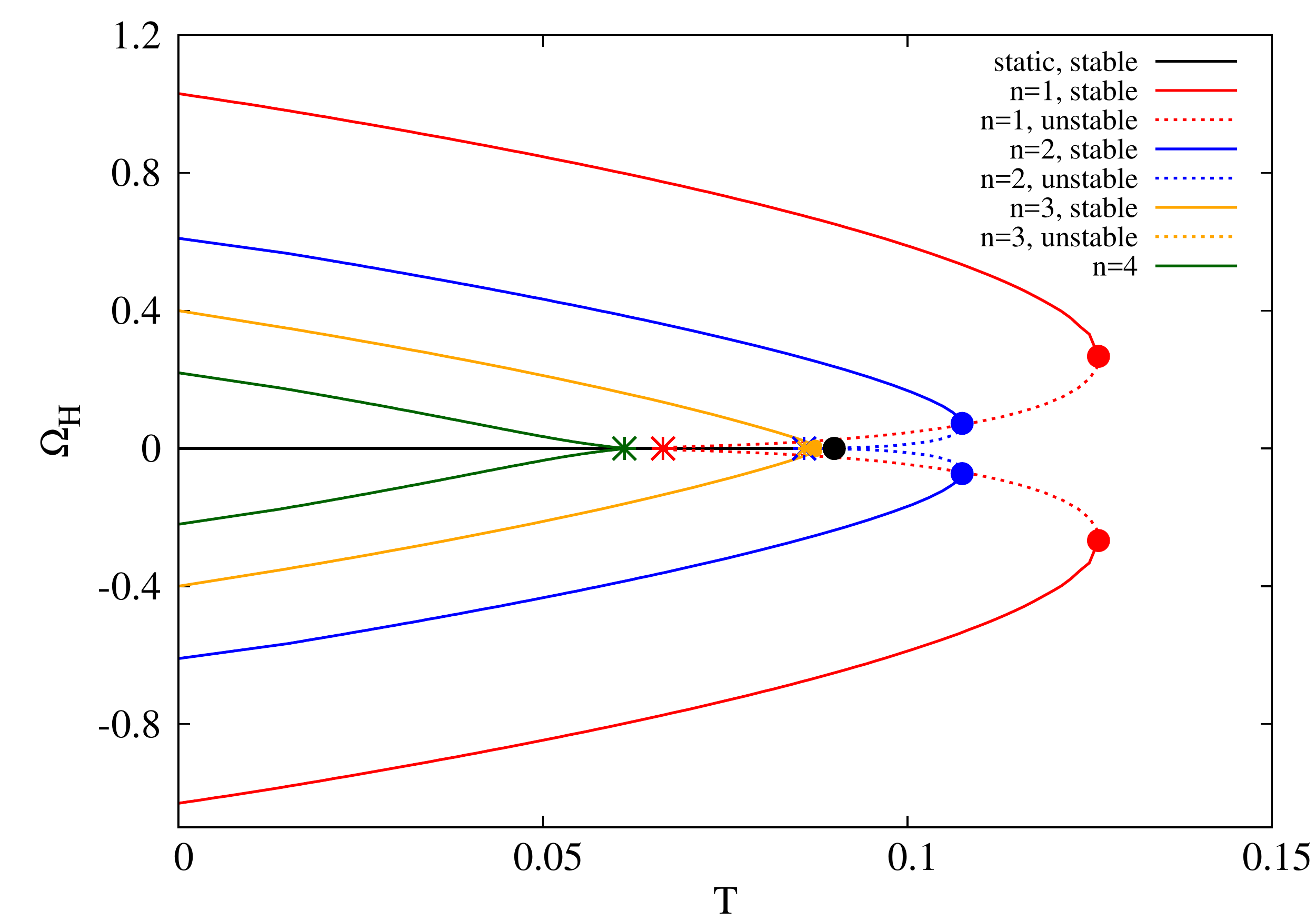}
         \caption{}
      \end{subfigure}
\vspace{-12pt}
     \caption{Angular velocity $\Omega_H$ versus temperature $T$ for $J=0$ black holes with $Q=-2.72$ and $L=10$. In (\textbf{a}), we show the $\lambda=5$ case, and in (\textbf{b}) $\lambda=10$. RN-AdS black holes have $\Omega_H=0$, and we mark them with a black solid line. The non-static $J=0$ black holes are marked in color lines, representing canonical ensembles with different excitation number $n$: $n=1$ in red, $n=2$ in blue, $n=3$ in orange, and $n=4$ in green. We do not show higher excitations, which appear closer and closer to $T=0$. Solid lines represent thermodynamically stable branches, and dashed lines the unstable ones. The dots represent the critical solution where the branch changes stability. The stars mark the points where the non-static $J=0$ black holes merge with the static solution.}

         \label{f5}

\end{figure}

In Figure \ref{f5}a,b, we show the angular velocity     $\Omega_H$ versus the temperature for $J=0$ black holes with $Q=-2.72$ and $L=10$, for $\lambda=5,10,$ respectively. The extremal excited black holes have non-vanishing angular velocity (see Figure \ref{f2}a in the previous section), and their angular velocity decreases with the number $n$. 

\textls[-15]{Concerning the non-extremal solutions, we can see that the angular velocity decreases monotonically with the temperature along the stable branches,} but it increases with the temperature along the unstable ones.~Note that each ensemble of excited black holes merges at a different temperature with the static ensemble (marked with a colored star).\enlargethispage{0.5cm}

\begin{figure}[H]
       \begin{subfigure}[b]{0.45\textwidth}
          \centering
     \includegraphics[width=52mm,scale=0.5,angle=-90]{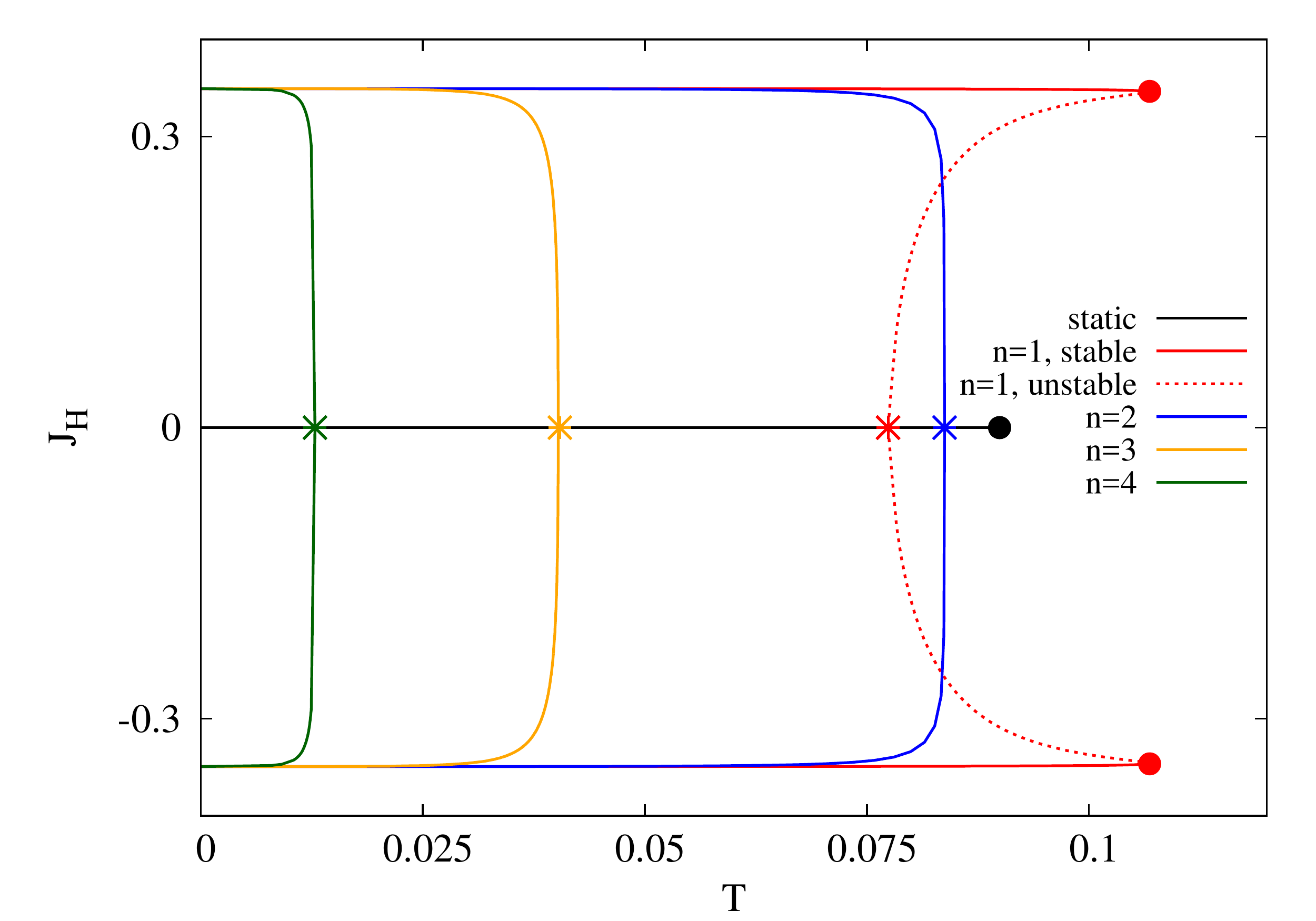}
         \caption{}
     \end{subfigure}
     \begin{subfigure}[b]{0.45\textwidth}
        \centering
     \includegraphics[width=52mm,scale=0.5,angle=-90]{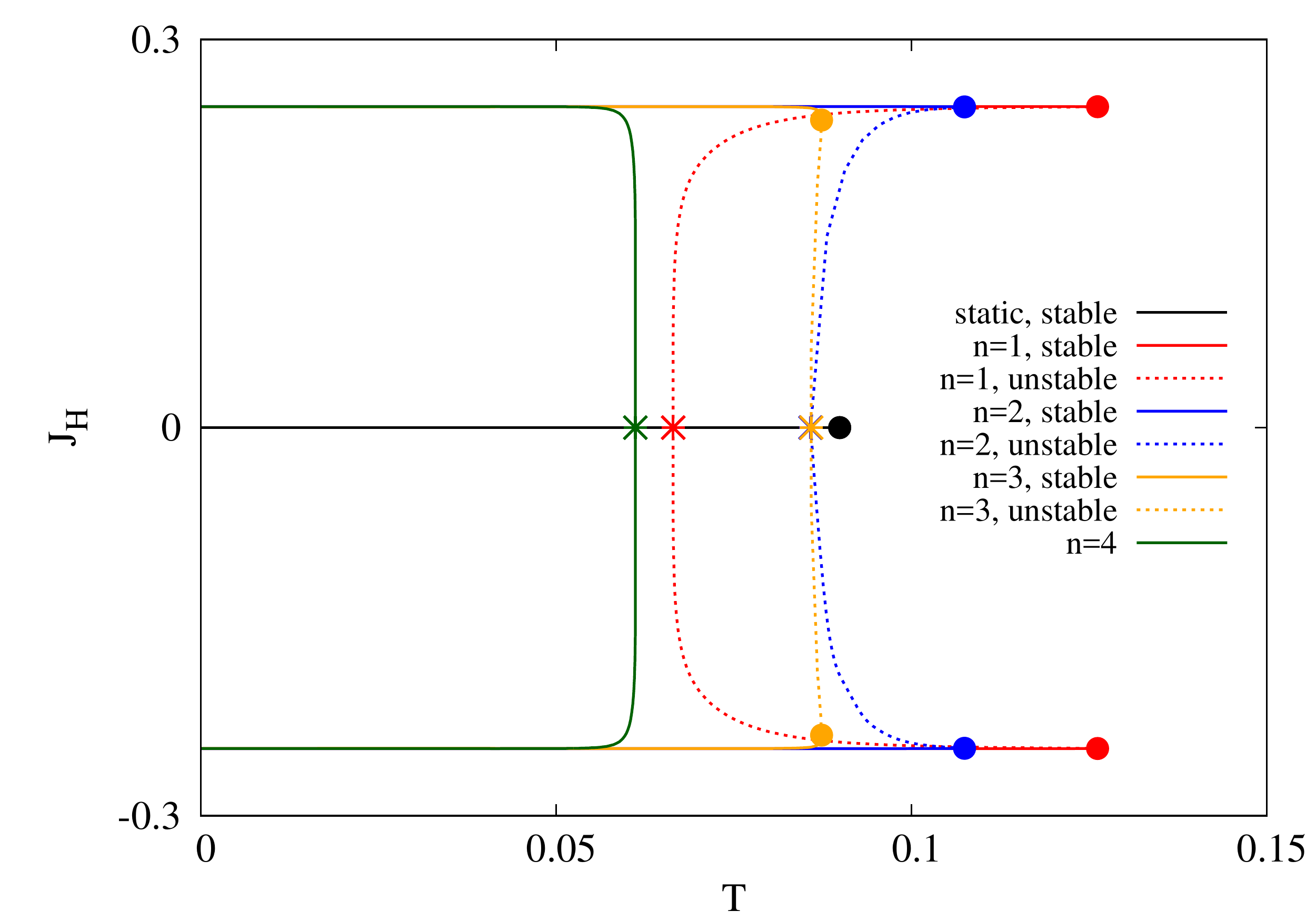}
         \caption{}
        \end{subfigure}
\vspace{-12pt}

     \caption{Horizon angular momentum $J_H$ vs. temperature $T$ for $J=0$ black holes with $Q=-2.72$, $L=10$ for the first four excitations. In (\textbf{a}), we show the $\lambda=5$ case, and in (\textbf{b}) $\lambda=10$. In both figures, the black line shows the RN-AdS black holes. For the non-static $J=0$ black holes, each color represents a canonical ensemble with different excitation number $n$: $n=1$ in red, $n=2$ in blue, $n=3$ in orange, and $n=4$ in green. Thermodynamically stable branches are marked with solid lines and unstable branches with dashed lines. The dots represent the critical solutions where stability changes. The stars mark the points where the non-static $J=0$ black holes merge with the static solution.}

         \label{f6}

\end{figure}

Another quantity of interest is the horizon angular momentum $J_H$, which we show in Figure~\ref{f6}a,b versus the temperature for $Q=-2.72$ and $L=10$, and $\lambda=5,10$, respectively. Surprisingly, the~horizon angular momentum along the stable branches does not change much with respect to the value of the extremal solution, and it drops to zero only when the temperature is very close to the critical point where the branch merges with the static ensemble. On the other hand, unstable branches exhibit a very different behavior, with the angular momentum being much more sensitive to the change of~temperature.

\begin{figure}[H]
      \begin{subfigure}[b]{0.45\textwidth}
          \centering
     \includegraphics[width=50mm,scale=0.5,angle=-90]{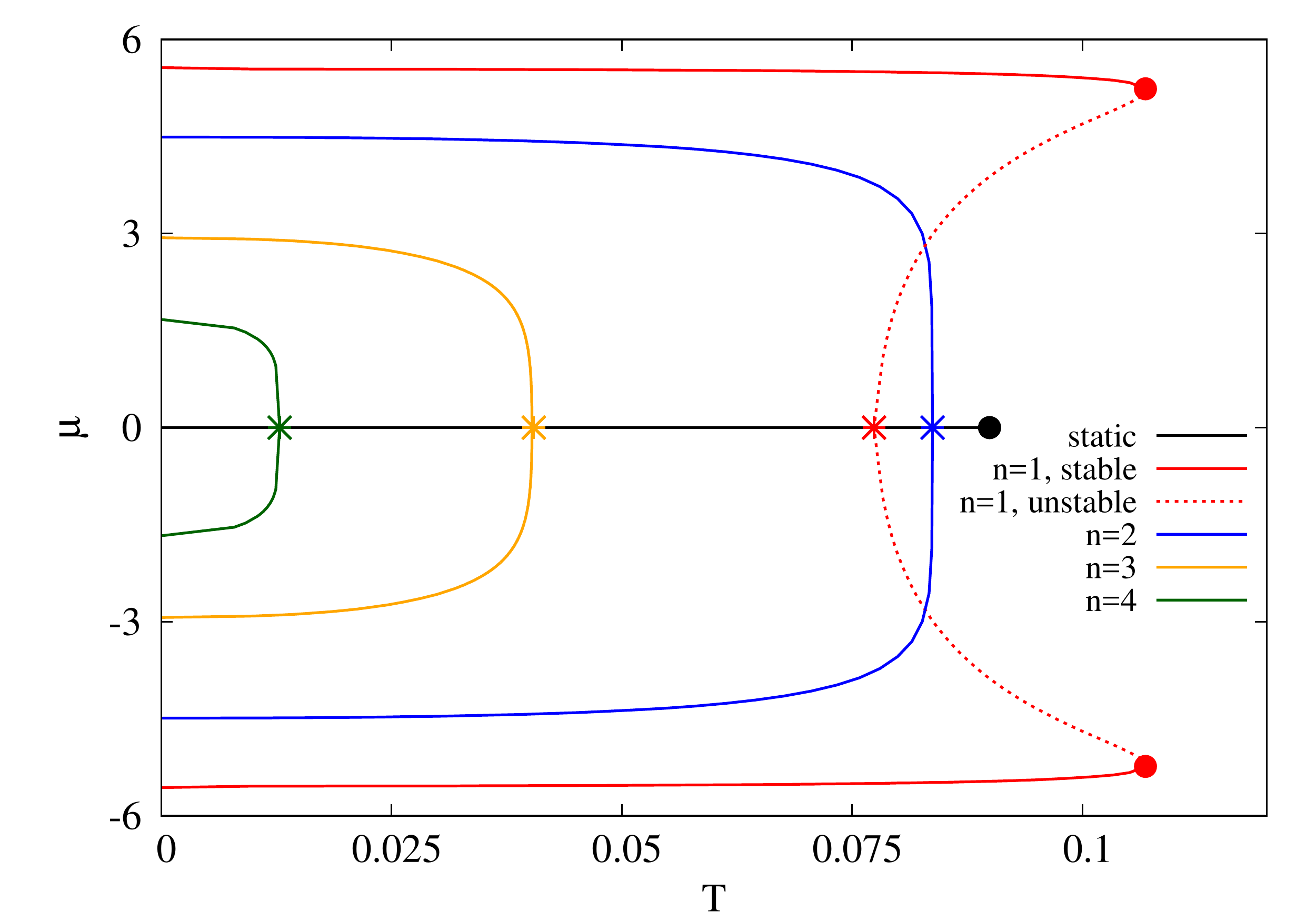}
         \caption{}
     \end{subfigure}
     \begin{subfigure}[b]{0.45\textwidth}
         \centering
     \includegraphics[width=50mm,scale=0.5,angle=-90]{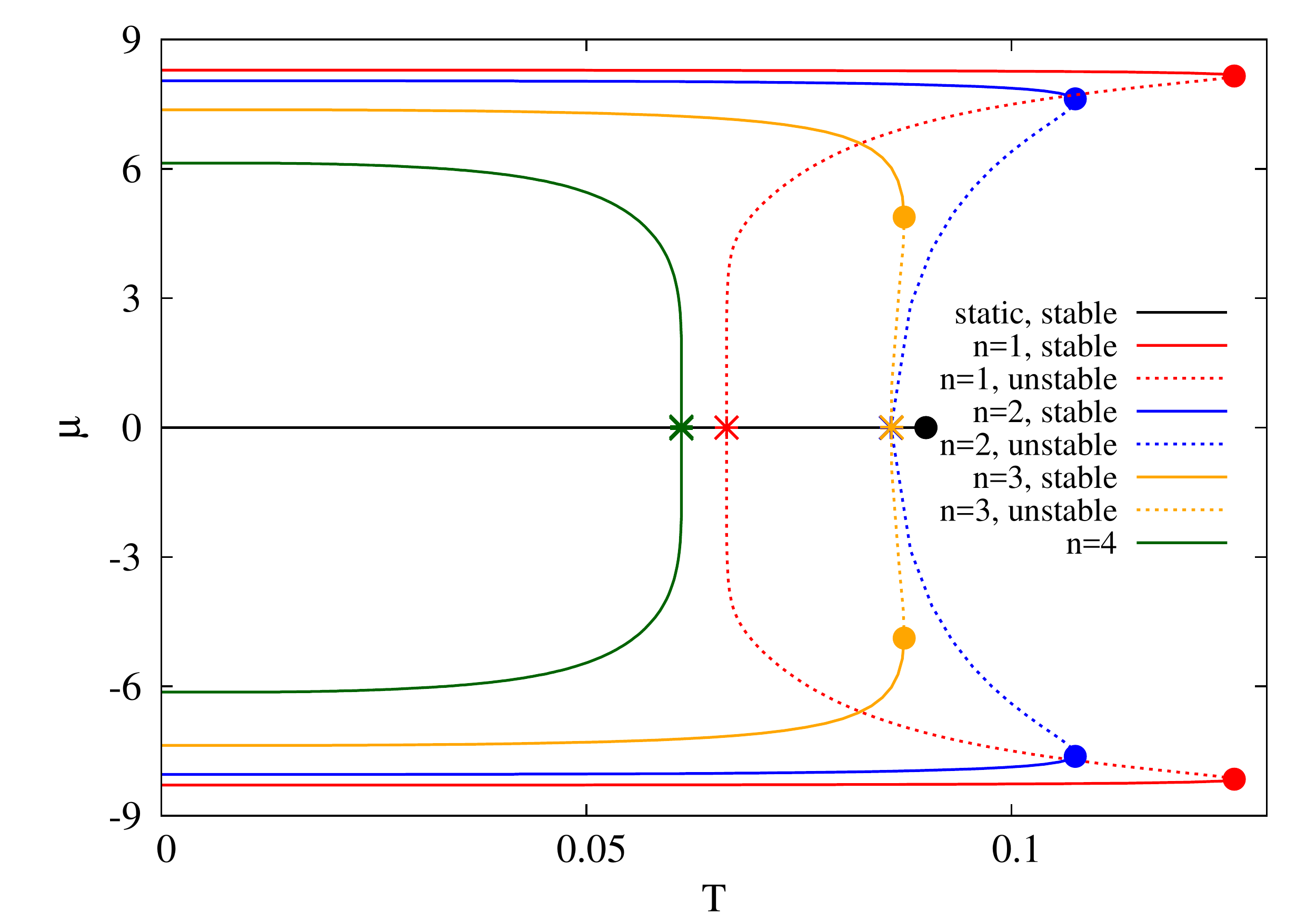}
         \caption{}
        \end{subfigure}
\vspace{-12pt}

     \caption{Magnetic moment $\mu$ versus temperature $T$ for $J=0$ black holes with $Q=-2.72$ and $L=10$ for the first four excitations. In (\textbf{a}), we show the $\lambda=5$ case, and in (\textbf{b}) $\lambda=10$. In both figures, the black solid line shows the RN-AdS black holes. For the non-static $J=0$ black holes, each color represents a~canonical ensemble with different excitation number $n$: $n=1$ in red, $n=2$ in blue, $n=3$ in orange, and $n=4$ in green. Solid lines represent thermodynamically stable branches, and dashed lines the unstable ones. Dots represent the critical solution where the branch changes stability. The stars mark the points where the non-static $J=0$ black holes merge with the static solution.}

         \label{f7}

\end{figure}

A similar situation can be observed in Figure \ref{f7}a,b, where we plot the magnetic moment $\mu$ versus the temperature for the same set of parameters and $\lambda=5,10$, respectively. It presents a behavior similar to $J_H$, the magnetic moment being almost constant along the stable branches, and only changing when the ensemble of excited black holes approaches the merging point with the static ensemble, or~along the unstable branches.

So far in this discussion, we have restricted the analysis to local thermodynamic stability, by~looking at the behavior of the heat capacity. However, global thermodynamic stability requires the moment of inertia to be positive $\varepsilon>0$. The numerical analysis reveals that the sign of $\varepsilon$ depends on the charges of the black hole, the branch number, and the CS coupling. For instance, configurations that are locally stable can be globally unstable. We can easily see an example of this situation in the $n=1$ non-static extremal $J=0$ configuration for $\lambda=5$. As we said, this configuration is locally stable in Figure \ref{f4}a, since $C_p>0$. However, the moment of inertia of the $n=1$ configuration is negative, as it can be seen in Figure \ref{f2}a: the slope of the red curve around the $n=1$ solution is negative, meaning $\varepsilon<0$. However, the rest of branches with $n>1$ have positive $\varepsilon$, meaning that these excited $T=0$ black holes with $n>1$ are in fact globally stable.

%%%%%%%%%%%%%%%%%%%%%%%%%%%%%%%%%%%%%%%%%%%%%%%

\vspace{-12pt}

%%%%%%%%%%%%%%%%%%%%%%%%%%%%%%%%%%%%%%%

\subsubsection{The Case for Non-Static $\Omega_H=0$ Black Holes}

The branch structure presented in Section \ref{sect3-1} includes another class of excited non-static extremal solutions, in this case with $\Omega_H=0$ and fixed values of the electric charge. From Figure \ref{f2}a, we can see that the magnitude of the total angular momentum of these configurations decreases with the excitation number. Hence, we cannot construct a canonical ensemble with such configurations, but we can identify these sets with grand canonical ensembles. We include in this section a brief discussion of their properties, in order to compare them with the non-static $J=0$ black holes.

\begin{figure}[H]
     \centering

     \begin{subfigure}[b]{0.45\textwidth}
     \includegraphics[width=50mm,scale=0.5,angle=-90]{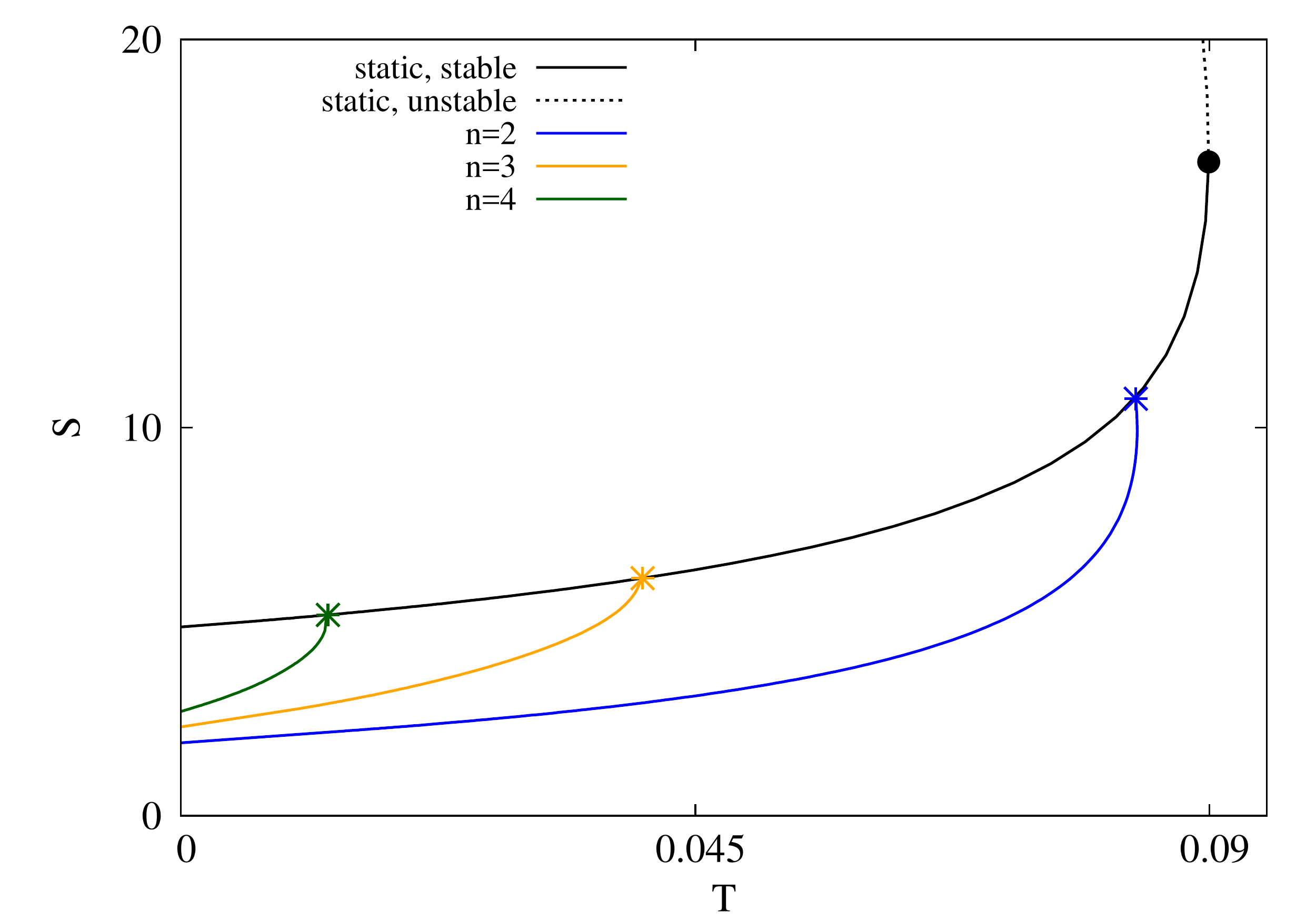}
         \caption{}
            \end{subfigure}
     \begin{subfigure}[b]{0.45\textwidth}
     \includegraphics[width=50mm,scale=0.5,angle=-90]{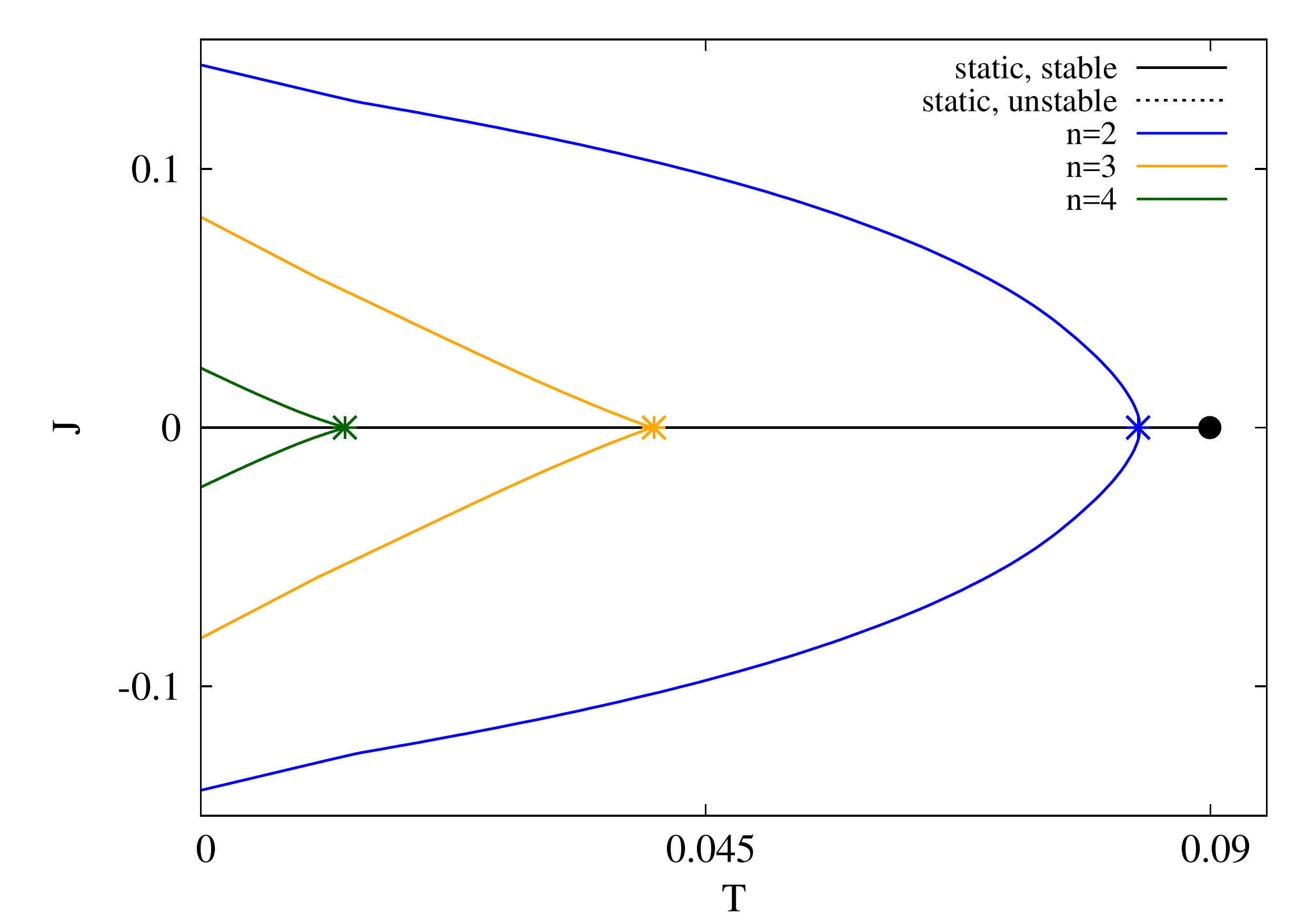}
         \caption{}
     \end{subfigure}
\vspace{-12pt}

     \caption{(\textbf{a}) Entropy $S$ and (\textbf{b}) total angular momentum $J$ versus temperature $T$ for $\Omega_H=0$, $Q=-2.72$, $L=10$ and $\lambda=5$ black holes. In black, we plot the RN-AdS black holes, in solid line stable, in dashed line unstable. Each branch of non-static $\Omega=0$ is colored according to its excitation number $n$: in blue, orange and green $n=2, 3$ and $4$, respectively.}
         \label{f8}

\end{figure}

In Figure \ref{f8}a, we show the entropy $S$ versus the temperature $T$ for the $\Omega_H=0$ black holes. We~choose $Q=-2.72$, $L=10$ and $\lambda=5$, so these solutions can, for instance, be compared with the ones in Figure \ref{f4}a. In black, we show the RN-AdS black hole, with its two phases (stable with a solid line, and unstable with a dashed line). Note that, in Figure \ref{f2}a, we can see that the first non-static $\Omega_H=0$ black hole solution is always found in the $n=2$ branch. In Figure \ref{f8}a, we plot the non-extremal $n=2$ black holes with a blue line. In orange and green, we show the $n=3$ and $n=4$ excited black holes as well.

%%%%%%%%%%%%%%%%%%%%%%%%%%%%%%%%%%%%%%%%%%%%%%%
%%%%%%%%%%%%%%%%%%%%%%%%%%%%%%%%%%%%%%%
For these ensembles, we can study local thermodynamic stability using the specific heat, calculated by taking variations of the entropy with respect to the temperature for constant $Q$ and $\Omega_H$ \cite{Chamblin:1999hg}: 
\begin{equation}
C_p = T \frac{\partial S}{\partial T}|_{\Omega_H,Q}.
\end{equation}

The branch structure is then very similar to the one we found for the $J=0$ excited configurations: 
for the lower modes, the temperature reaches a maximum and eventually the branch merges with the static solutions at some point (for instance, in Figure \ref{f8}a, the blue $n=2$ curve has a maximum temperature a bit before merging with the black curve). The branches of higher excitations on the other hand reach the maximum temperature when they merge with the static solutions. This reveals that, for~lower excitations, these ensembles can present a locally unstable branch, while, for higher excitations, they are locally stable. In particular, the numerical results suggest that the extremal $\Omega_H=0$ solutions possess local thermodynamic stability independently of the value of the charges.

Note that the non-static $\Omega_H=0$ extremal black holes do not share the same near-horizon geometry, as it happens in the non-static $J=0$ extremal solutions. This is the reason why, at $T=0$, each excited $\Omega_H=0$ black hole possesses a different entropy, which increases towards the static value as the number $n$ increases.

\textls[-15]{In Figure \ref{f8}b, we show the total angular momentum $J$ of these configurations versus the temperature. The maximum angular momentum is reached at $T=0$, and it decreases as the temperature rises.} Hence, we can see here that, as the temperature increases and black holes in the branches get closer to the static set, the solutions become continuously more and more similar to the static RN-AdS black hole.

\begin{figure}[H]
     \centering

     \begin{subfigure}[b]{0.45\textwidth}
     \includegraphics[width=52mm,scale=0.5,angle=-90]{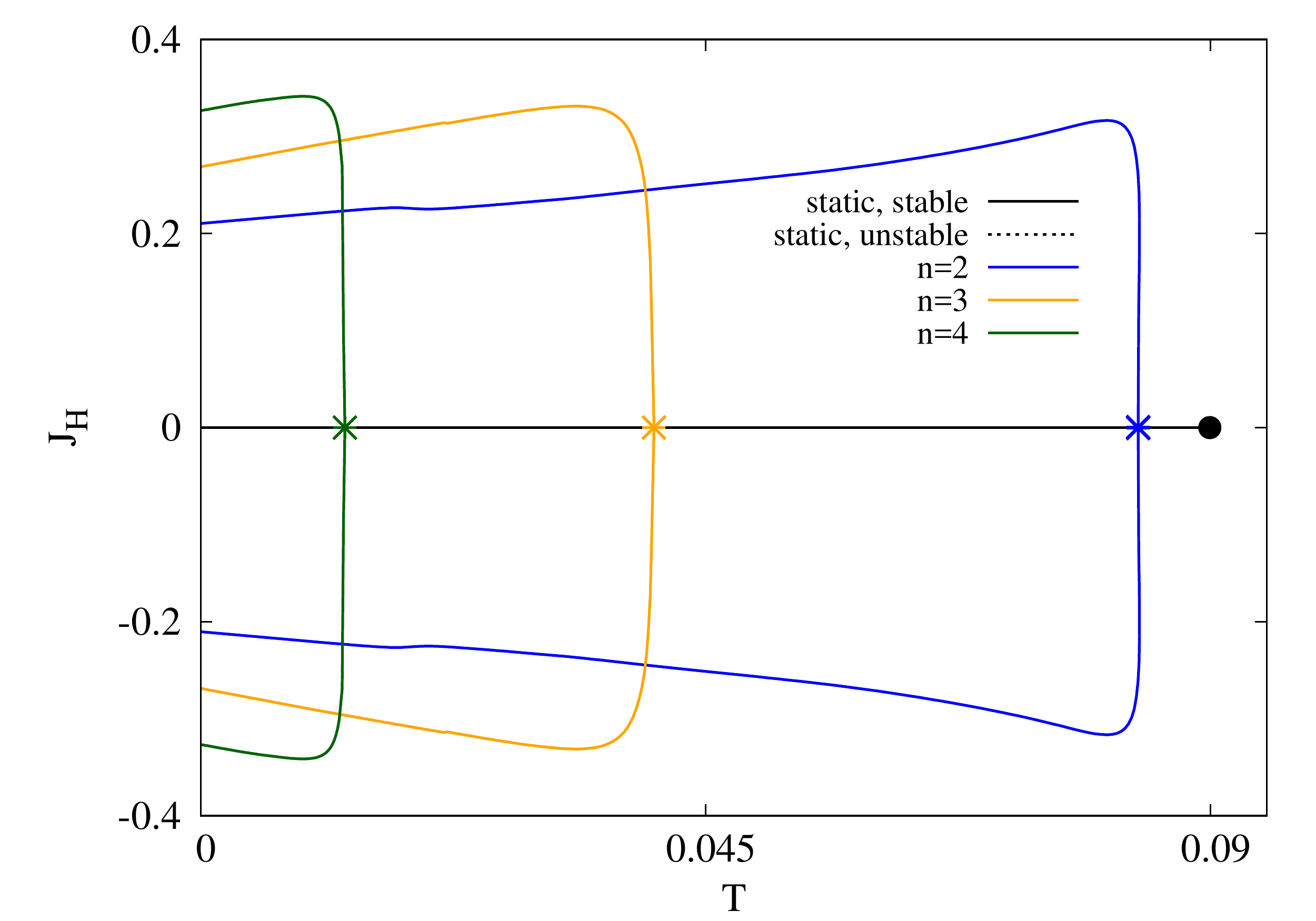}
         \caption{}
           \end{subfigure}
     \begin{subfigure}[b]{0.45\textwidth}
     \includegraphics[width=52mm,scale=0.5,angle=-90]{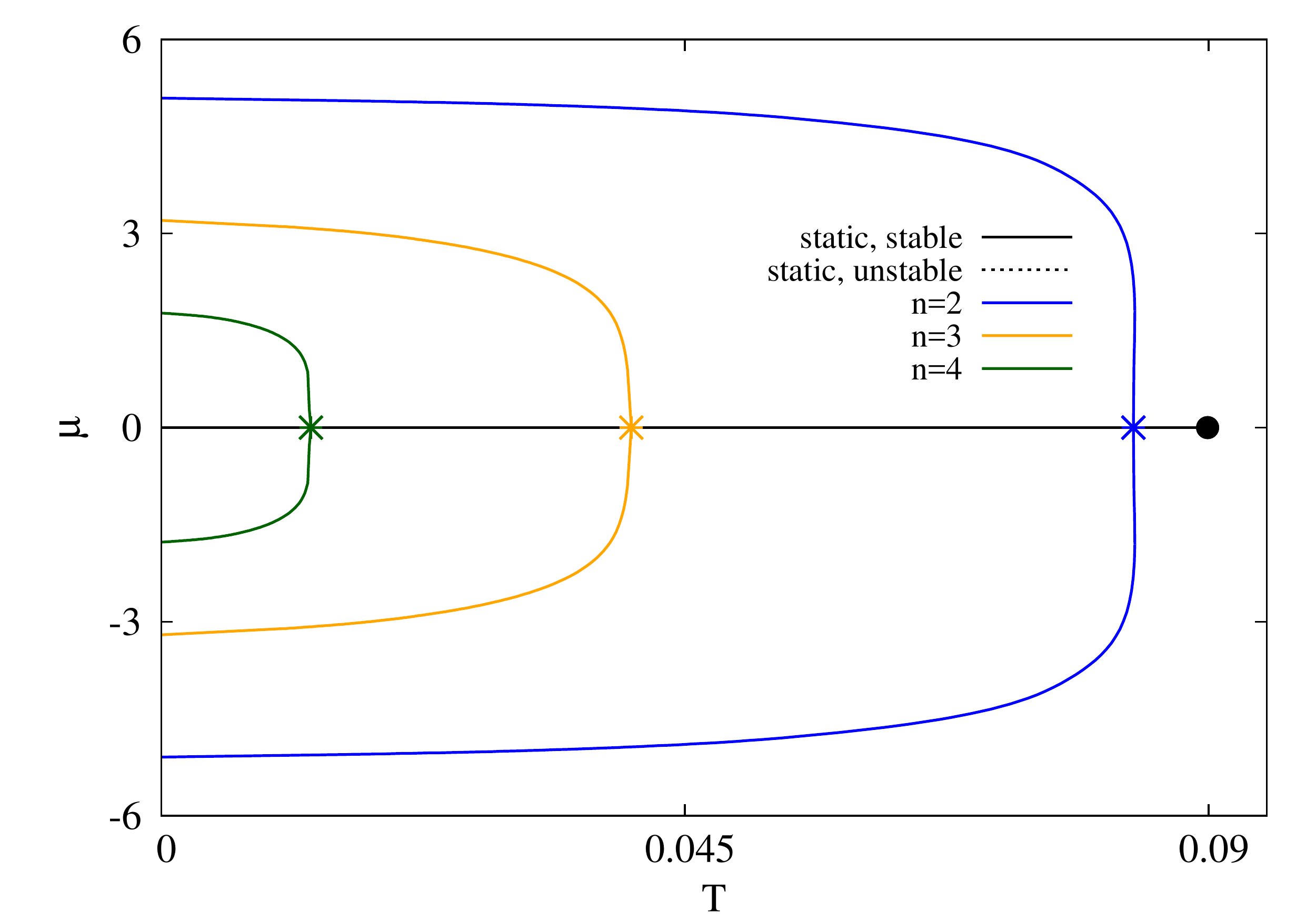}
         \caption{}
     \end{subfigure}
\vspace{-12pt}

     \caption{(\textbf{a}) horizon angular momentum $J_H$ and (\textbf{b}) magnetic moment $\mu$ versus temperature $T$ for $\Omega_H=0$, $Q=-2.72$, $L=10$ and $\lambda=5$ black holes. In black, we plot the RN-AdS black holes, solid line is stable, dashed line is unstable. Each branch of non-static $\Omega=0$ is colored according to its excitation number $n$: in blue, orange and green $n=2, 3$ and $4$, respectively.}

         \label{f9}

\end{figure}

Although the horizon angular velocity is zero, the configuration is storing angular momentum behind the horizon. This can be seen in Figure \ref{f9}a, where we show the horizon angular momentum $J_H$ for the same parameters as before. Here, note that the maximum horizon angular momentum is not reached at $T=0$, but at some non-extremal configuration, which differs depending on each excitation number. In particular, note that, for the $n=2$ branch, the maximum $|J_H|$ is very close to the maximum~temperature. 

In Figure \ref{f9}b, we show the magnetic moment $\mu$ as a function of the temperature. The magnetic moment behaves similarly to the total angular momentum. The maximum magnetic moment is reached at the $T=0$ solutions, but decreases with the excitation number. 

%%%%%%%%%%%%%%%%%%%%%%%%%%%%%%%%%%%%%%%%%%%%%%%
%%%%%%%%%%%%%%%%%%%%%%%%%%%%%%%%%%%%%%%

%%%%%%%%%%%%%%%%%%%%%%%%%%%%%%%%%%%%%%%%%%
\section{Conclusions} \label{s4}

In this paper, we have studied 5-dimensional charged and rotating black holes in Einstein--Maxwell--Chern--Simons theory with negative cosmological constant. We have considered the case in which the Chern--Simons coupling $\lambda > 2$, and both angular momenta are of equal magnitude. We have seen that, in this case, charged and rotating extremal black holes possess a complicated branch structure. These new branches are found in between the extremal RN-AdS black hole and the lowest excitation ($n=1$) of a family of non-static $J=0$ black holes. The branches contain an intricate structure of cusps, branching points and radially excited solutions. Each branch can be labeled with an integer number $n$, which is related to the number of nodes of the non-static $J=0$ solutions inside the branch. The mass of each branch increases with the number $n$. The black holes in these branches present very unique features, like non-uniqueness with respect to the global charges, degeneracy of the horizon geometry, and isolation of the extremal RN-AdS solution from the rotating and charged ones. Concerning this last point, we have seen that the  extremal RN-AdS solution cannot be reached by decreasing continuously the angular momentum of a rotating and charged extremal black hole, but, instead, it is approached asymptotically as the excitation number $n$ is increased to infinity. 

All of these features of the extremal black holes are qualitatively very similar to what was previously found in the asymptotically flat case, with these configurations being essentially an AdS generalization of the previously known black holes. The non-static $J=0$ black holes can now be understood as different excitations of the same near-horizon geometry, which matches physically different solutions at the bulk, with different values of the charges like the mass and magnetic moment at the AdS$_5$ boundary.

We have also made a study of the thermal properties of the non-static $J=0$ black holes. In~particular, we have considered canonical ensembles of configurations with constant electric charge, $J=0$, and different excitation number $n$. We have studied their thermodynamic stability and branch structure as a function of $T$, finding that the thermodynamic stability of each branch depends on the specific values of the electric charge and Chern--Simons coupling. While lower excitation numbers possess thermodynamically unstable black holes, for a large enough excitation number $n$, it is possible to obtain a branch of stable configurations. In addition, we have studied the set of non-static $\Omega_H=0$ solutions, which present similar features to their $J=0$ counterparts. Finally, we have seen that these families of excited black holes are not present for arbitrary large values of the temperature as in the static case, but instead they merge with a non-extremal static black hole for large enough values of $T$~and $S$.

As an outlook, we would like to mention that the solutions we have discussed here may also be present in the more general case of non-equal angular momenta. In this case, the numerical analysis is more involved, since the configurations are described by partial differential equations. On the other hand, it has been recently shown that the CPL black hole can be generalized to include a non-trivial magnetization of the AdS boundary \cite{Blazquez-Salcedo:2017cqm}. It should be possible to magnetize in a similar way the excited black holes that we have discussed here.

%%%%%%%%%%%%%%%%%%%%%%%%%%%%%%%%%%%%%%%%%%
\vspace{6pt} 

%%%%%%%%%%%%%%%%%%%%%%%%%%%%%%%%%%%%%%%%%%
\acknowledgments{\textls[-5]{J.L.B.-S. would like to thank Jutta Kunz, Francisco Navarro-L\'erida, and Eugen Radu for
helpful }comments and discussions.
J.L.B.-S. gratefully acknowledges support by
the Deutsche Forschungsgemeinschaft Research Training Group 1620 ``Models of Gravity'', and Grant FP7, 
Marie Curie Actions, People, International Research Staff Exchange Scheme (IRSES-606096).}

%%%%%%%%%%%%%%%%%%%%%%%%%%%%%%%%%%%%%%%%%%
\conflictsofinterest{The author declares no conflict of interest.} 

%%%%%%%%%%%%%%%%%%%%%%%%%%%%%%%%%%%%%%%%%%

\reftitle{{References}}


\begin{thebibliography}{999}

%%%%%%%%%%%%%%%%%%%%%%%%%%%%%%%%%%%%%%%%%%%%%%%%%%%%%%%%%%%%%%%%%%%%%%%%%%%%%%
%\cite{Witten:1998qj}
\bibitem{Witten:1998qj}
Witten, E.~Anti de Sitter space and holography.
\emph{Adv. Theor. Math. Phys.} \textbf{1998}, {\em 2},  253--291.
 
\bibitem{Maldacena:1997re}
Maldacena, J.M. The Large N limit of superconformal field theories and supergravity.
\emph{Adv. Theor. Math. Phys.}  \textbf{1998}, {\em 2}, 231--252.





\bibitem{Hawking:1998kw}
Hawking, S.W.; Hunter, C.J.; Taylor-Robinson, M.M. Rotation and the AdS-CFT correspondence.
  \emph{Phys.~Rev.~D} \textbf{1999}, {\em 59}, 064005.

  

\bibitem{Mitra:1999ge} 
  Mitra, P. Thermodynamics of charged anti-de Sitter black holes in canonical ensemble.
 \emph{ Phys. Lett. B} \textbf{1999}, {\em 459}, 119--124.

  

\bibitem{Chamblin:1999tk} 
  Chamblin, A.; Emparan, R.; Johnson, C.V.; Myers, R.C. Charged AdS black holes and catastrophic holography.
  \emph{Phys. Rev. D} \textbf{1999}, {\em 60}, 064018.


\bibitem{Chamblin:1999hg} 
Chamblin, A.; Emparan, R.; Johnson, C.V.; Myers, R.C. Holography, thermodynamics, and fluctuations of charged AdS black holes.
  \emph{Phys. Rev. D} \textbf{1999}, {\em 60}, 104026.


\bibitem{Blazquez-Salcedo:2016vwa} 
 Bl\'azquez-Salcedo,  J.L.; Kunz, J.; Navarro-L\'erida, F.;  Radu, E. Static Einstein--Maxwell magnetic solitons and black holes in an odd dimensional AdS spacetime.
  \emph{Entropy} \textbf{2016}, {\em 18}, 438. 
 
  

\bibitem{Kunz:2007jq}
  Kunz, J.; Navarro-L\'erida, F.; Radu, E. Higher dimensional rotating black holes in Einstein--Maxwell theory with negative cosmological constant.
  \emph{Phys. Rev. B} \textbf{2007}, {\em 649},  463--471.

\bibitem{Blazquez-Salcedo:2016rkj} 
 Bl\'azquez-Salcedo,  J.L.; Kunz, J.; Navarro-L\'erida, F.;  Radu, E. Charged rotating black holes in Einstein-- Maxwell--Chern--Simons theory with a negative cosmological constant.
  \emph{Phys. Rev. D} \textbf{2017}, {\em 95},  064018.

\bibitem{Figueras:2014dta}
  Figueras, P.;~Tunyasuvunakool,  S.~Black rings in global anti-de Sitter space.
  \emph{J. High Energy Phys.} \textbf{2015},    doi:10.1007/JHEP03(2015)149.

 \bibitem{Cvetic:2004hs}
 Cvetic,  M.; Lu, H.;  Pope, C.N. Charged Kerr-de Sitter black holes in five dimensions.
  \emph{Phys. Rev. B} \textbf{2004}, {\em 598},  273--278.

\bibitem{Gutowski:2004ez}
 Gutowski, J.B.; Reall, H.S. Supersymmetric AdS5 black holes.
 \emph{ J. High Energy Phys.} \textbf{2004}, {\em 2}, 6.





\bibitem{Chong:2005hr}
Chong, Z.W.; Cvetic, M.; Lu, H.; Pope, C.N. General nonextremal rotating black holes in minimal five-dimensional gauged supergravity.
 \emph{Phys. Rev. Lett.} \textbf{2005},  {\em95},  161301.


\bibitem{Chong:2006zx}
 Chong, Z.W.; Cvetic, M.; Lu, H.; Pope, C.N. Non-extremal rotating black holes in five-dimensional gauged supergravity.
  \emph{Phys. Lett. B} \textbf{2007}, {\em 644},  192--197.



\bibitem{Chong:2005da}
  Chong, Z.W.; Cvetic, M.; Lu, H.; Pope, C.N. Five-dimensional gauged supergravity black holes with independent rotation parameters.
  \emph{Phys. Rev. D} \textbf{2005}, {\em 72},  041901.


\bibitem{Cvetic:2004ny}
 Cvetic, M.; Lu, H.; Pope, C.N. Charged rotating black holes in five dimensional $U$(1) 3 gauged $N =$ 2 supergravity.
  \emph{Phys. Rev. D}  \textbf{2004}, {\em 70}, 081502.

\bibitem{Blazquez-Salcedo:2017cqm} 
 Bl\'azquez-Salcedo,  J.L.; Kunz, J.; Navarro-L\'erida, F.;  Radu, E. AdS 5 magnetized solutions in minimal gauged supergravity.
  \emph{Phys. Rev. B} \textbf{2007}, {\em 771}, 52--58.


\bibitem{Grunau:2015ana} 
  Grunau, S.;  Neumann, H. Thermodynamics of a rotating black hole in minimal five-dimensional gauged supergravity.
   \emph{Class. Quantum Gravity} \textbf{ 2015}, {\em 32},  175004.

  
  

\bibitem{Mir:2016dio} 
{Mir, M.; Mann,  R.B. Charged rotating AdS black holes with Chern--Simons coupling.
  \emph{Phys. Rev. D} \textbf{2017}, {\em 95},~024005.}

 
	

\bibitem{Blazquez-Salcedo:2013muz} 
 Bl\'azquez-Salcedo,  J.L.; Kunz, J.; Navarro-L\'erida, F.;  Radu, E. Sequences of extremal radially excited rotating black holes.  \emph{Phys. Rev. Lett.} \textbf{2014}, {\em 112},  011101.


\bibitem{Blazquez-Salcedo:2015kja}
 Bl\'azquez-Salcedo,  J.L.; Kunz, J.; Navarro-L\'erida, F.;  Radu, E. Radially excited rotating black holes in Einstein--Maxwell--Chern--Simons theory.
  \emph{Phys. Rev. D} \textbf{2015}, {\em 92},  044025.

\bibitem{Ashtekar:1984zz}
 Ashtekar, A.; Magnon,  A. Asymptotically anti-de Sitter space-times.
 \emph{ Class. Quant. Grav. } \textbf{1984}, {\em 1},  L39.


\bibitem{Ashtekar:1999jx} 
   Ashtekar, A.; Das, S. Asymptotically anti-de Sitter spacetimes: Conserved quantities.
    \emph{Class. Quant. Grav.} \textbf{2000},  {\em 17}, L17.

\bibitem{Balasubramanian:1999re}
Balasubramanian, V.; Kraus, P. A stress tensor for anti-de Sitter gravity.
\emph{Commun. Math. Phys.} \textbf{1999}, \emph{208}, 413--428.  


\bibitem{Kastor:2009wy} 
  Kastor, D.; Ray, S.; Traschen, J. Enthalpy and the mechanics of AdS black holes.
 \emph{ Class. Quantum Gravity}  \textbf{2009}, {\em 26},~195011.



\bibitem{Cvetic:2010jb} 
  Cvetic, M.; Gibbons, G.W.; Kubiznak, D.; Pope, C.N. Black hole enthalpy and an entropy inequality for the thermodynamic volume.
    \emph{Phys. Rev. D} \textbf{2011}, {\em 84}, 024037.

  

\bibitem{Dolan:2013ft} 
Dolan, B.P.; Kastor, D.; Kubiznak, D.; Mann, R.B.; Traschen, J. Thermodynamic volumes and isoperimetric inequalities for de Sitter black holes.
  \emph{Phys. Rev. D} \textbf{2013}, {\em 87}, 104017.

  

\bibitem{Altamirano:2014tva} 
Altamirano, N.; Kubiznak, D.; Mann, R.B.; Sherkatghanad, Z. Thermodynamics of rotating black holes and black rings: Phase transitions and thermodynamic volume.
  \emph{Galaxies} \textbf{2014}, {\em 2}, 89--159.


  

\bibitem{COLSYS1}
Ascher, U.; Christiansen, J.; Russell, R.D. A collocation solver for mixed order systems of boundary value problems.
 \emph{Math. Comput. } \textbf{1979}, {\em 33}, 659--679.



\bibitem{COLSYS2}
Ascher, U.; Christiansen, J.; Russell, R.D. Collocation software for boundary-value ODEs.
 \emph{ACM Trans.} \textbf{1981}, {\em 7},  209--222.	  


\bibitem{Sen:2005wa} 
  Sen, A. Black hole entropy function and the attractor mechanism in higher derivative gravity.
  \emph{J. High Energy~Phys.} \textbf{2005},  {\em 9}, 038.

  

\bibitem{Astefanesei:2006dd} 
  Astefanesei, D.; Goldstein, K.; Jena, R.P.; Sen, A.; Trivedi, S.P. Rotating attractors.
  \emph{J. High Energy Phys. }  \textbf{2006}, {\em 10}, 058.


  

\bibitem{Goldstein:2007km} 
Goldstein, K.; Jena, R.P. One entropy function to rule them all…
  \emph{J. High Energy Phys.} \textbf{2007}, {\em 11}, 049.



\bibitem{Kunduri:2007qy}
  Kunduri, H.K.; Lucietti, J. Near-horizon geometries of supersymmetric AdS5 black holes.
  \emph{J. High Energy~Phys.} \textbf{2007}, {\em12},  015.
  
  
  

\bibitem{Cai:2014oca} 
  Cai, R.G.; Yang, R.Q. Paramagnetism-ferromagnetism phase transition in a dyonic black hole.
  \emph{Phys. Rev. D} \textbf{2014}, {\em 90},  081901.



\bibitem{Caldarelli:1999xj} 
 Caldarelli, M.M.; Cognola, G.; Klemm, D. Thermodynamics of Kerr-Newman-AdS black holes and conformal field theories.
  \emph{Class. Quantum Gravity} \textbf{2000}, {\em 17}, 399.



\bibitem{Dolan:2013yca} 
  Dolan, B.P. On the thermodynamic stability of rotating black holes in higher dimensions--a comparison of thermodynamic ensembles.
  \emph{Class. Quantum Gravity} \textbf{2014}, {\em 31}, 135012.

 \end{thebibliography}
\end{document}